\documentclass[useAMS,usenatbib]{mn2e}
\usepackage{subfigure}
\usepackage{graphicx}

\title[Are luminous radio-loud AGN triggered by galaxy interactions?]{Are luminous radio-loud active galactic nuclei triggered by galaxy 
interactions?}
\author[C. Ramos Almeida et al.]
{\parbox{\textwidth}{C. Ramos Almeida$^{1}$\thanks{E-mail:C.Ramos@sheffield.ac.uk},
P.~S.~Bessiere$^{1}$,
C.~N.~Tadhunter$^{1}$,
P.~G.~P\' erez-Gonz\'alez$^{2,3}$,
G.~Barro$^{2}$,
K.~J.~Inskip$^{4}$,
R.~Morganti$^{5,6}$,
J.~Holt$^{7}$, \&
D. Dicken$^{8}$}\vspace{0.4cm}\\
\parbox{\textwidth}{$^{1}$Department of Physics and Astronomy, University of Sheffield, Sheffield, S3 7RH, UK\\
$^{2}$Departamento de Astrof\' isica, Facultad de CC. F\' isicas, Universidad Complutense de Madrid, E-28040 Madrid, Spain\\
$^{3}$Associate Astronomer at Steward Observatory, The University of Arizona\\
$^{4}$Max-Planck-Institut f\"ur Astronomie, K\"oningstuhl 17, D-69117 Heidelberg, Germany\\
$^{5}$Netherlands Institute for Radio Astronomy, Postbus 2, 7990 AA Dwingeloo, the Netherlands\\
$^{6}$Kapteyn Astronomical Institute, University of Groningen, Postbus 800, 9700 AV Groningen, the Netherlands\\
$^{7}$Leiden Observatory, Leiden University, PO Box 9513, 2300 RA Leiden, the Netherlands\\
$^{8}$Department of Physics and Astronomy, Rochester Institute of Technology, 84 Lomb Memorial Drive, Rochester NY 14623, USA}}

\begin{document}

\date{}

\pagerange{\pageref{firstpage}--\pageref{lastpage}} \pubyear{2010}

\maketitle

\label{firstpage}

\begin{abstract}
We present the results of a comparison between the optical morphologies of a complete sample of 46 southern 
2Jy radio galaxies at intermediate redshifts ($0.05<z<0.7$) and those of two control samples of 
quiescent early-type galaxies: 55 ellipticals 
at redshifts $z\leq0.01$ from the Observations of Bright Ellipticals at Yale (OBEY) survey, 
and 107 early-type galaxies at redshifts $0.2<z<0.7$ in the Extended Groth Strip (EGS).
Based on these comparisons, we discuss the role of galaxy interactions in the triggering 
of powerful radio galaxies (PRGs). We find that a significant fraction of quiescent ellipticals 
at low and intermediate redshifts show evidence for disturbed morphologies at relatively high  
surface brightness levels, which are likely the result of past or on-going galaxy interactions. However, the morphological 
features detected in the galaxy hosts of the PRGs (e.g. tidal tails, shells, bridges, etc.) are up to 2 magnitudes
brighter than those present in their quiescent counterparts. Indeed, if we consider the same surface brightness
limits, the fraction of disturbed morphologies is considerably smaller in the quiescent population 
(53\% at $z<0.2$ and 48\% at $0.2\leq z<0.7$) than in the PRGs (93\% at $z<0.2$ and 95\% at $0.2\leq z<0.7$ 
considering strong-line radio galaxies only). This supports a scenario in which PRGs represent a fleeting 
active phase of a subset of the elliptical galaxies that have recently undergone mergers/interactions.  
However, we demonstrate that only a small proportion ($\la$20\%) of disturbed early-type galaxies are capable of hosting 
powerful radio sources. 


\end{abstract}


\section{Introduction}
\label{intro}

Simulations of hierarchical galaxy evolution predict that the periods of black hole (BH) growth and nuclear 
activity are intimately tied to the growth of the host galaxy, and that the triggering of the main phase of 
this nuclear activity in gas-rich mergers will always be accompanied by a major galaxy-wide starburst 
\citep{Kauffmann00,diMatteo05,Springel05,Hopkins08a,Hopkins08b,Somerville08}. 
However, the order of events and the timescales involved in both the triggering of the merger-induced starburst and the 
nuclear activity remain uncertain (see e.g. \citealt{Canalizo00,Wild10,Tadhunter11}).

Based on cosmological simulations, \citet{Hopkins08b} suggested a bimodality in the BH  
triggering mechanisms: luminous quasar-like activity is associated with the formation of classical bulges
and ellipticals via galaxy mergers, whereas the less luminous Seyfert-like activity is associated with
the formation of pseudobulges and bulgeless galaxies via secular processes. 
Under the assumption that major, gas-rich mergers trigger quasar activity, 
\citet{Hopkins08b} reproduce the observed quasar luminosity function from z=0 to z=6. 
They also compare with a secular model in which the nuclear activity is driven by bars 
or instabilities and show that, although these processes
probably dominate at luminosities typical of Seyfert galaxies, they contribute very little to the 
$z\ga1$ quasar luminosity density. 

From an observational point of view, a bimodality in formation mechanisms 
(and hence in AGN triggering) is supported by the fact that, whereas classical bulges and elliptical galaxies  
follow a close correlation between velocity dispersion and BH mass 
\citep{Kormendy95,Magorrian98,Gebhardt00,Ferrarese00,Greene06}, bulgeless galaxies and 
those with pseudobulges show no clear evidence for such a correlation \citep{Kormendy11}.  


Galaxy interactions are one of the most efficient mechanism to transport the cold gas required to trigger and feed
AGN to the center of galaxies \citep{Kauffmann00,Cox06,Cox08,Croton06,diMatteo07} and many observational studies 
of powerful AGN (i.e. quasar-like) have revealed a high incidence of interaction signatures 
in their host galaxies (\citealt{Heckman86,Hutchings87,Smith89,Canalizo01,Canalizo07,Bennert08}, Ramos Almeida et al. 2011). 
However, all these studies lack comparisons with appropriate samples of quiescent (i.e. non-active) galaxies to 
confirm that the percentage of interacting systems in powerful AGN is larger than in the
quiescent population. Indeed, based on high spatial resolution Hubble Space Telescope (HST) images of a sample of nearby
radio galaxies and quasars, \citet{Dunlop03} found evidences that their hosts are indistinguishable from quiescent 
ellipticals of similar mass. 
Moreover, for moderately luminous AGN (i.e. Seyfert galaxies, L$_{bol}\sim 10^{42}-10^{45}$ erg~s$^{-1}$) several 
studies find that the incidence of disturbed morphologies is not significantly enhanced over the 
general population (e.g. \citealt{Malkan98,Grogin05,Georgakakis09,Gabor09}; Cisternas et al. 2011), 
although some others find the opposite result (e.g. \citealt{Keel96,Kuo08,Koss10,Ellison11}).


In our previous work (Ramos Almeida et al. 2011; hereafter \citealt{Ramos11}) we studied the optical 
morphologies of a complete sample of 46 southern 2Jy radio galaxies at intermediate redshifts ($0.05<z<0.7$)
and found that the overall majority of the sample (up to 85\%) show peculiarities in their 
optical morphologies at relatively high levels of surface brightness. Our study indicates that galaxy 
interactions are likely to play a key role in the triggering of AGN/jet activity, especially in the 
case of strong-line radio galaxies (SLRGs)\footnote{SLRGs comprise narrow-line radio galaxies (NLRGs), 
broad-line radio galaxies (BLRGs) and quasars, i.e.
they are radio galaxies with strong and high equivalent width emission lines.}, of which 94\% appear disturbed. 
On the other hand, of the weak-line
radio galaxies (WLRGs)\footnote{WLRGs have optical spectra dominated by the stellar continua of the host galaxies and small 
emission line equivalent widths (EW$_{[O III]}<10$ \AA; \citealt{Tadhunter98}).} in the 2Jy sample, 
only 27\% show clear evidence for tidal features. 
These results are consistent with the most accepted explanation for the differences between
the properties of SLRGs and WLRGs, in which SLRGs are powered by cold gas accretion, while WLRGs are fuelled by
accretion of hot gas from their X-ray coronae \citep{Allen06,Best06,Hardcastle07,Balmaverde08,Buttiglione10}. 

The high percentage of disturbed morphologies in the 2Jy sample of radio galaxies contrasts with 
the results found for lower luminosity AGN in, for example, the recent extensive study by Cisternas et al. 
(2011; hereafter \citealt{Cisternas11}).
However, the \citealt{Ramos11} and \citealt{Cisternas11} studies can be reconciled by considering 
the differences in the depth of the observations and 
sample selection. First, the images employed in \citealt{Cisternas11} were obtained with the 
Advanced Camera for Surveys (ACS) on the HST and are not as deep as our Gemini Multi-Object Spectrograph South 
(GMOS-S)/Gemini observations: the limiting surface brightness level of those HST images is 23.3 mag~arcsec$^{-2}$ 
in the ACS F814W filter, whereas in \citealt{Ramos11} we detected features as faint as 
26.3 mag~arcsec$^{-2}$ in the same band (using colour transformations for elliptical galaxies from 
\citealt{Fukugita95}). Thus, the imaging observations used in \citealt{Cisternas11},
with a surface brightness limit 3 magnitudes brighter than ours, are not sensitive enough to reveal faint diffuse tidal features in 
their AGN and control samples, even if present.
Second, the AGN in \citealt{Cisternas11} were selected at X-ray wavelengths and the galaxy hosts are mostly  
disks with luminosities more typical of luminous Seyfert galaxies (median L$_{bol}\sim 10^{44.8} erg~s^{-1}$). 
In contrast, the AGN in \citealt{Ramos11} were selected according to their radio 
emission and the majority have quasar-like luminosities that are typically an order of magnitude higher than those 
of the \citealt{Cisternas11} sample (median value of L$_{bol}\sim 10^{45.7} erg~s^{-1}$ for the 
SLRGs in the 2Jy sample\footnote{L$_{bol}$ was derived
from the [O III] luminosities of the individual galaxies listed in \citet{Dicken09} by applying the 
bolometric correction factor of 3500 reported in \citet{Heckman04} for low-redshift quasars.}) and they are 
almost exclusively hosted by elliptical galaxies. 
Thus, the differences between the findings of \citealt{Ramos11} and \citealt{Cisternas11}
can be explained by the luminosity-dependent bimodality in the BH triggering mechanisms suggested by \citet{Hopkins08b} and 
\citet{Kormendy11}, as well as by the differences in depth between the observations employed. 


If galaxy interactions are the main triggering mechanism for radio-loud AGN activity in our sample, 
then it is expected that the signs of 
morphological disturbance will be stronger and more common in the radio source host galaxies than in the general population 
of quiescent ellipticals. 
Studies of nearby red galaxies (e.g. \citealt{vanDokkum05}) have shown that the majority of quiescent 
luminous ellipticals were assembled through gas-poor mergers, which explains their old stellar populations
and high central densities. However, triggering and feeding a powerful radio source (and occasionally star 
formation; see \citealt{Tadhunter11}) is likely to require a larger amount of cold gas 
to be accreted into the central regions of the galaxy. 
The morphological signatures
of gas-rich interactions (such as tidal tails, shells, bridges, etc.) are brighter 
than those produced in gas-poor interactions \citep{Naab06,Bell06,McIntosh08}. Thus, the surface brightness 
of the various morphological features can be used as an indicator of the type of interaction. 
In addition, the features resulting from gas-rich interactions are expected to be visible over time-scales between 0.5 and 1.5 
Gyr \citep{Fevre00,Patton02,Conselice03,Kawata06}, whereas those formed in gas-poor interactions are visible 
for only $\sim$150 Myr \citep{Bell06}. 

This is the second in a series of papers based on the analysis of the optical morphologies of PRGs. In the first
paper (\citealt{Ramos11}), we presented deep Gemini images for the 2Jy sample and 
compared the results found for the PRGs with various samples of quiescent 
ellipticals and/or red galaxies from the literature \citep{Malin83,vanDokkum05,Tal09}. However, 
only the observations reported in the study of \citet{Malin83}, which is based 
on photographic plates, have similar surface brightness depth to our PRG sample ($\mu_V\la 25.5~mag~arcsec^{-2}$). 
After comparing with the latter study we concluded that 
the percentage of morphological disturbance of the PRGs (up to 85\%) greatly exceeds that found for quiescent 
elliptical galaxies when the same surface brightness depth is considered ($\sim$10\%). However, in order to make a more quantitative 
comparison, it is necessary to develop control samples 
of elliptical galaxies at similar redshifts and masses, probing the same scales and depths, and using CCD imaging data. 
In this paper we present the results from such a 
comparison. In Section 2 we describe the control sample selection and observations. In Section 3 we present the 
observational results. The comparison between the morphologies of PRGs and quiescent elliptical galaxies
is discussed in Section 4, and the main conclusions from this work are summarized in Section 5. Throughout this paper we 
assume a cosmology with H$_0$ = 70 km s$^{-1}$ Mpc$^{-1}$, $\Omega_m$ = 0.27, and $\Omega_{\Lambda}$ =0.73.

\section{Sample Selection and observations}
\label{selection}

The objects studied in \citealt{Ramos11} comprise all powerful radio galaxies 
(PRGs) and quasars from the \citet{Tadhunter93} sample of 2Jy radio galaxies with S$_{2.7GHz}\ge$ 
2.0 Jy, steep radio spectra $\alpha_{2.7}^{4.8} > 0.5~(F_{\nu}\propto\nu^{-\alpha})$, 
declinations $\delta<+10\degr$ and redshifts $0.05<z<0.7$ (see Table 1 in \citealt{Ramos11}). 
It is itself a subset of the \citet{Wall85} complete sample of 2Jy radio sources.
The z $>$ 0.05 limit ensures that the radio galaxies are genuinely powerful sources, while the z $<$ 0.7 limit ensures
that sources are sufficiently nearby for detailed morphological studies.

In terms of the optical classification, the sample comprises 24\% WLRGs and 76\% SLRGs \citep{Tadhunter98}.
Considering the radio morphologies, Fanaroff-Riley II (FRII) sources constitute the majority of the sample (72\%),
13\% are Fanaroff-Riley I (FRI), and the remaining 15\% are compact, steep-spectrum (CSS) or 
Gigahertz-peaked spectrum (GPS) sources (see Table 1 in \citealt{Ramos11}).

Moderately luminous AGN (e.g. those studied in \citealt{Cisternas11}) have a relatively high surface density 
and can be easily selected in deep field surveys, together with appropiate control samples of quiescent galaxies. 
On the contrary, quasars and radio galaxies are much rarer and cannot be studied using narrow, deep field surveys. 
In consequence, it is more challenging to develop control samples for such objects.  

Since radio galaxies are almost invariably associated with elliptical hosts (see e.g. \citealt{Heckman86} and 
\citealt{Dunlop03}), we searched in the literature for 
samples of elliptical galaxies with similar masses and redshifts as our 2Jy PRGs. In addition, similar angular resolutions
and depths are required to probe the same spatial scales and surface brightness levels. Our sample of 46 PRGs was
imaged with the Gemini Multi-Object Spectrograph South (GMOS-S) on the 8.1-m Gemini South telescope at Cerro Pach\'on
under good seeing conditions (median seeing of 0.8\arcsec, ranging from 0.4\arcsec to 1.15\arcsec). 
The GMOS-S detector \citep{Hook04} comprises three adjacent CCDs, giving a field-of-view (FOV) of 5.5$\times$5.5 arcmin$^2$, with a pixel 
size of 0.146\arcsec.
The morphological features reported in \citealt{Ramos11} have a median surface brightness of 
$\mu_V = 23.6~mag~arcsec^{-2}$ and $\Delta\mu_V \sim [21, 26]~mag~arcsec^{-2}$. 
See \citealt{Ramos11} for a more detailed description of the GMOS-S observations.  
Thus, after considering all these factors, we finally selected control samples of elliptical 
galaxies in two redshift ranges which best match the 2Jy sample host galaxies: the Observations of Bright Ellipticals at Yale (OBEY) survey and 
the Extended Groth Strip (EGS) sample.

\subsection{The OBEY survey} 
\label{obey}

The OBEY survey \citep{Tal09} is a volume-limited and statistically complete 
sample of 55 luminous elliptical galaxies selected from the Nearby Galaxies 
Catalog (\citealt{Tully88}; see Table \ref{data}). 
It consists of all elliptical galaxies in the \citet{Tully88} catalog with declinations between -85 and +10, 
at distances from 16 to 40 Mpc, and M$_B<$-20.4 mag, once corrected to the cosmology considered here. 
The sample comprises galaxies from a wide range of different environments: 36\% are field galaxies, 33\% are in groups, 
and 18\% are in clusters, including members of the Virgo, Fornax, Centaurus, and Antlia clusters. 
These galaxy environments were determined from the literature by \citet{Tal09} and no 
classification is reported in that study for the remaining 13\%. This wide variety of environments in the OBEY survey 
matches those typical of FRII radio galaxies, which are found in field/groups as well as in moderately rich clusters 
\citep{Prestage88,Smith90,Zirbel97} and constitute the majority of our PRG sample (72\%).

\begin{table*}
\centering
\begin{tabular}{lccccccccc}
\hline
\hline
GALAXY     & R.A.            &      Dec      &  z        &   M$_B$  & (B-V)   &  Obs. date  &  Tc         &  Morphology  & Group    \\
\hline
NGC 0584   &   01:31:20.7    &    -06:52:05  &	0.0060   &   -20.98  &  0.92  & 2006-08-22   &  0.076	   & 2S,B     &  1  \\
NGC 0596   &   01:32:52.1    &    -07:01:55  &	0.0062   &   -20.52  &  0.90  & 2006-08-23   &  0.110	   & S,F      &  2  \\  
NGC 0720   &   01:53:00.5    &    -13:44:19  &	0.0058   &   -20.68  &  0.96  & 2006-08-24   &  0.079	   & 2F       &  2  \\  	 
NGC 1199   &   03:03:38.4    &    -15:36:49  &	0.0085   &   -20.39  &  0.97  & 2006-01-25   &  0.067	   & \dots    &  5  \\  
NGC 1209   &   03:06:03.0    &    -15:36:41  &	0.0086   &   -20.66  &  0.95  & 2006-01-26   &  0.116	   & T,3F,[D] &  2  \\ 
NGC 1399   &   03:38:29.1    &    -35:27:03  &	0.0047   &   -20.64  &  0.98  & 2008-09-30   &  0.064	   & \dots    &  5  \\
NGC 1395   &   03:38:29.8    &    -23:01:40  &	0.0057   &   -20.59  &  0.94  & 2006-01-27   &  0.094	   & 3S       &  2  \\ 
NGC 1407   &   03:40:11.9    &    -18:34:49  &	0.0059   &   -21.32  &  0.93  & 2006-01-28   &  0.083	   & \dots    &  5  \\
NGC 2865   &   09:23:30.2    &    -23:09:41  &	0.0087   &   -21.00  &  0.78  & 2006-03-28   &  0.193	   & 3S,2T,[D]&  2  \\
NGC 2974   &   09:42:33.3    &    -03:41:57  &	0.0064   &   -20.93  &  0.95  & 2006-02-05   &  0.110	   & S,[D]    &  2  \\        
NGC 2986   &   09:44:16.0    &    -21:16:41  &	0.0076   &   -21.00  &  0.99  & 2006-01-30   &  0.045	   & [B]      &  5  \\
NGC 3078   &   09:58:24.6    &    -26:55:37  &	0.0086   &   -20.93  &  0.97  & 2006-01-31   &  0.103	   & \dots    &  5  \\
NGC 3258   &   10:28:53.6    &    -35:36:20  &	0.0093   &   -20.62  &  0.92  & 2006-02-01   &  0.123	   & [2N]     &  5  \\         
NGC 3268   &   10:30:00.6    &    -35:19:32  &	0.0093   &   -20.58  &  0.96  & 2006-02-03   &  0.087	   & S        &  2  \\
NGC 3557B  &   11:09:32.1    &    -37:20:59  &	0.0096   &   -20.43  &  0.86  & 2006-02-04   &  0.182	   & 2I       &  2  \\
NGC 3557   &   11:09:57.6    &    -37:32:21  &	0.0103   &   -22.48  &  0.87  & 2006-01-31   &  0.111	   & F,[S]    &  2  \\
NGC 3585   &   11:13:17.1    &    -26:45:18  &	0.0047   &   -21.23  &  0.91  & 2006-03-29   &  0.048	   & 2S       &  2  \\
NGC 3640   &   11:21:06.8    &    +03:14:05  &	0.0041   &   -21.08  &  0.92  & 2006-04-01   &  0.142	   & S,4F     &  2  \\
NGC 3706   &   11:29:44.4    &    -36:23:29  &	0.0099   &   -21.44  &  0.93  & 2006-01-26   &  0.120	   & 2S       &  2  \\
NGC 3904   &   11:49:13.2    &    -29:16:36  &	0.0052   &   -20.41  &  0.94  & 2006-01-28   &  0.108	   & S        &  2  \\
NGC 3923   &   11:51:01.8    &    -28:48:22  &	0.0058   &   -21.53  &  0.95  & 2006-01-29   &  0.100	   & 4S       &  2  \\  		       
NGC 3962   &   11:54:40.1    &    -13:58:30  &	0.0060   &   -21.03  &  0.95  & 2006-03-30   &  0.059	   & S        &  2  \\  		     
NGC 4105   &   12:06:40.8    &    -29:45:37  &	0.0064   &   -20.66  &  0.87  & 2006-02-02   &  0.109	   & 2F,T     &  1  \\
NGC 4261   &   12:19:23.2    &    +05:49:31  &	0.0074   &   -21.71  &  0.98  & 2006-04-03   &  0.053	   & T,F      &  2  \\
NGC 4365   &   12:24:28.2    &    +07:19:03  &	0.0041   &   -20.82  &  0.97  & 2006-03-29   &  0.070	   & F        &  2  \\
IC 3370    &   12:27:37.3    &    -39:20:16  &	0.0097   &   -21.53  &  0.89  & 2006-04-02   &  0.192	   & F,S,D    &  2  \\
NGC 4472   &   12:29:46.7    &    +08:00:02  &	0.0033   &   -22.12  &  0.97  & 2008-06-09   &  0.000	   & \dots    &  5  \\      
NGC 4636   &   12:42:49.9    &    +02:41:16  &	0.0031   &   -20.98  &  0.93  & 2006-03-28   &  0.066	   & F        &  2  \\
NGC 4645   &   12:44:10.0    &    -41:45:00  &	0.0087   &   -21.18  &  0.95  & 2009-04-18   &  0.000	   & \dots    &  5  \\
NGC 4697   &   12:48:35.9    &    -05:48:03  &	0.0041   &   -21.97  &  0.92  & 2006-04-04   &  0.091	   & \dots    &  5  \\
NGC 4696   &   12:48:49.3    &    -41:18:40  &	0.0098   &   -22.35  &  0.94  & 2008-06-08   &  0.075	   & S,D      &  2  \\
NGC 4767   &   12:53:52.9    &    -39:42:52  &	0.0099   &   -21.43  &  0.93  & 2008-06-10   &  0.000	   & 2S,[D]   &  2  \\
NGC 5011   &   13:12:51.8    &    -43:05:46  &	0.0105   &   -21.40  &  0.89  & 2006-04-05   &  0.077	   & \dots    &  5  \\
NGC 5018   &   13:13:01.0    &    -19:31:05  &	0.0093   &   -21.76  &  0.85  & 2008-06-03   &  0.184	   & 3T,3S,[D]&  2  \\
NGC 5044   &   13:15:24.0    &    -16:23:08  &	0.0092   &   -21.31  &  0.98  & 2008-06-06   &  0.041	   & \dots    &  5  \\        
NGC 5061   &   13:18:05.1    &    -26:50:14  &	0.0069   &   -21.49  &  0.85  & 2006-04-01   &  0.104	   & T,S      &  2  \\
NGC 5077   &   13:19:31.7    &    -12:39:25  &	0.0093   &   -20.82  &  0.98  & 2006-03-30   &  0.061	   & [S],[D]  &  5  \\
NGC 5576   &   14:21:03.7    &    +03:16:16  &	0.0049   &   -20.70  &  0.88  & 2008-06-06   &  0.122	   & 3T,S     &  2  \\
NGC 5638   &   14:29:40.4    &    +03:14:00  &	0.0055   &   -21.42  &  0.94  & 2008-06-07   &  0.036	   & T,S      &  2  \\
NGC 5812   &   15:00:55.7    &    -07:27:26  &	0.0065   &   -20.88  &  0.94  & 2008-06-08   &  0.080	   & T        &  2  \\
NGC 5813   &   15:01:11.2    &    +01:42:07  &	0.0065   &   -21.07  &  0.95  & 2008-06-09   &  0.054	   & \dots    &  5  \\
NGC 5846   &   15:06:29.3    &    +01:36:20  &	0.0057   &   -21.46  &  0.98  & 2008-06-07   &  0.068	   & 3S,2N    &  2,3\\
NGC 5898   &   15:18:13.5    &    -24:05:53  &	0.0070   &   -20.79  &  0.92  & 2006-04-05   &  0.114	   & 3T,D,2N  &  2,3\\
NGC 5903   &   15:18:36.5    &    -24:04:07  &	0.0085   &   -21.18  &  0.89  & 2006-04-05   &  0.075	   & \dots    &  5  \\  	
IC 4797    &   18:56:29.7    &    -54:18:21  &	0.0089   &   -21.05  &  0.92  & 2006-08-23   &  0.226	   & T,I,[D]  &  2  \\
IC 4889    &   19:45:15.1    &    -54:20:39  &	0.0085   &   -20.85  &  0.88  & 2006-08-18   &  0.158	   & F        &  2  \\
NGC 6861   &   20:07:19.5    &    -48:22:13  &	0.0094   &   -21.10  &  0.95  & 2006-08-19   &  0.123	   & D        &  4  \\
NGC 6868   &   20:09:54.1    &    -48:22:46  &	0.0095   &   -21.36  &  0.97  & 2006-08-20   &  0.096	   & \dots    &  5  \\
NGC 6958   &   20:48:42.6    &    -37:59:51  &	0.0090   &   -20.83  &  0.86  & 2006-08-22   &  0.122	   & 3S,[D]   &  2  \\
NGC 7029   &   21:11:52.0    &    -49:17:01  &	0.0094   &   -20.41  &  0.86  & 2006-08-22   &  0.085	   & \dots    &  5  \\
NGC 7144   &   21:52:42.4    &    -48:15:14  &	0.0064   &   -20.66  &  0.91  & 2006-08-17   &  0.100	   & \dots    &  5  \\
NGC 7196   &   22:05:54.8    &    -50:07:10  &	0.0097   &   -20.62  &  0.91  & 2006-08-19   &  0.171	   & S,[D]    &  2  \\
NGC 7192   &   22:06:50.1    &    -64:18:58  &	0.0099   &   -20.85  &  0.92  & 2006-08-20   &  0.096	   & S        &  2  \\  	
IC 1459    &   22:57:10.6    &    -36:27:44  &	0.0060   &   -20.88  &  0.96  & 2006-08-18   &  0.137	   & 4S       &  2  \\
NGC 7507   &   23:12:07.6    &    -28:32:23  &	0.0052   &   -20.51  &  0.94  & 2006-08-21   &  0.084	   & S        &  2  \\
\hline		     			      		 
\end{tabular}						 
\caption{Full classification of the OBEY survey ordered by R.A. Columns 2, 3, and 4 list R.A., declination and 
spectroscopic redshift as reported in the NASA/IPAC Extragalactic Database (NED).
Columns 5, 6, and 7 correspond to the B-band absolute magnitudes from \citet{Tully88} and corrected to H$_0=70~km~s^{-1}~Mpc^{-1}$, 
the Vega (B-V) colors within effective radius from \citet{Michard05} and \citet{deVaucouleurs91}, and the 
dates of observation. Column 8 lists the tidal parameter reported in \citet{Tal09}.
Columns 9 and 10 list our morphological classification (T: Tail; F: Fan; B: Bridge; S: Shell; D: Dust feature;  
2N: Double Nucleus; 3N: Triple Nucleus; A: Amorphous Halo; I: Irregular feature. Brackets indicate uncertain
identification of a feature), and division in groups: 1) galaxy pair or group in tidal interaction; 2) galaxies 
showing T,F,S,D,A,I; 3) multiple nuclei (inside a 10 kpc); 4) galaxies with dust as the only detected feature, 5) isolated galaxies with no sign
of interaction. Features with uncertain identification have not been considered in the statistics discussed in this study.}
\label{data}
\end{table*}

\begin{figure}
\centering
\includegraphics[width=8.5cm]{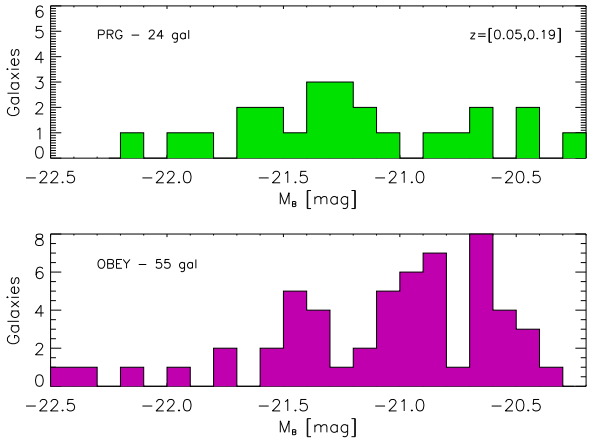}
\caption{Comparison between the B-band absolute magnitudes of the PRGs in the 2Jy sample at 
$z<0.2$ (top panel) and those of the \citet{Tal09} sample of quiescent ellipticals at 
z$\leq$0.01 (bottom panel).}
\label{tal1}
\end{figure}

Thus, we have a sample of 55 giant eliptical galaxies at redshifts z$\leq$0.01 and with absolute magnitudes M$_B$=[-22.5,-20.4] mag. 
If we assume no evolution for massive elliptical galaxies since z=0.2 \citep{Cimatti06,Faber07}, we can compare the 
OBEY sample with the PRGs in the 2Jy sample at $z<0.2$. In Figure \ref{tal1} we show a comparison between the absolute
magnitudes of the 24 PRGs with $z<0.2$ and the 55 quiescent ellipticals from
\citet{Tal09}. The M$_B$ values for the PRGs have been calculated from the Galactic extinction-, cosmological dimming-, 
and k-corrected r'-band magnitudes reported in \citealt{Ramos11}. 
Colours of elliptical galaxies from \citet{Fukugita95} have been used to convert 
the magnitudes to the B-band, resulting in absolute magnitudes within the interval M$_B=[-22.1,-20.3]$ mag. 
From the comparison between the two M$_B$ distributions shown in Figure \ref{tal1}, the 
significance level of the Kolmogorov-Smirnov (KS) statistic is 0.04 (i.e. there is only a 4\% chance that the two 
distributions are drawn from the same parent population). The low value of the KS probability 
is due to the larger number of M$_B>-21$ mag ellipticals in the OBEY sample compared to the PRGs. However,
both distributions span the same range in absolute magnitude and  
we prefer to keep the complete sample of 55 quiescent elliptical galaxies rather than reducing the number of 
fainter objects. In addition, as we discuss in Section \ref{classification}, we do not find any significant correlation 
between luminosity and the level of morphological disturbance.   


The 55 galaxies in the OBEY survey were imaged with 
Y4KCam, which is a 4Kx4K CCD camera optimzed for wide-field broad-band imaging mounted on the 1 m SMARTS 
telescope at the Cerro Tololo Inter-American Observatory (CTIO) between 2006 and 2009. 
The final images are a combination of 
several pointings of 300 s in the V-band, resulting in very deep frames with exposure times between 4200 and 
7200 s. \citet{Tal09} reported a detection threshold of $\mu_V \sim$ 27.7 mag (in this work we have
measured a surface brightness of $\mu_V$ = 28.2 mag~arcsec$^{-2}$ for the faintest feature detected, as  
shown in Section \ref{mu}).

The data were binned in order to improve the signal-to-noise of the images, which have a 
final pixel size of 1.156\arcsec~and a typical value of the seeing of $\sim$1.7\arcsec.
More details on the sample selection and observations can be found in \citet{Tal09} and are 
summarized in Table \ref{data}. The images are then deeper than the GMOS-S images
of the PRGs in the 2Jy sample, for which the faintest detected feature has a $\mu_V$ = 26.2 mag~arcsec$^{-2}$, 
and both the pixel size and the average seeing are larger than those of the 2Jy radio galaxies
(0.146\arcsec/pixel and FWHM$\sim$0.8\arcsec~respectively for the GMOS-S images). 
However, considering that the galaxies in the OBEY sample are at a median distance of 36 Mpc 
(spatial scale of $\sim$170 pc~arcsec$^{-1}$) and the PRGs (those at $z<0.2$) at 423 Mpc 
(spatial scale of $\sim$1700 pc~arcsec$^{-1}$) the effective resolution  will be better in the case
of the OBEY survey images. Thus, considering that the latter images 
are deeper and the resolution better, we will likely detect fainter and smaller features than 
for the PRGs. Even in the case of large-scale diffuse structures, the fact that the OBEY 
images are two magnitudes deeper than those of the 2Jy sample will allow us to detect them. 
Summarising, by using the same classification technique employed for the PRGs, 
we will be able to detect the same morphological signatures, if present, in the OBEY survey. 
Any possible bias will lead to a relative enhancement of the number of detected features
in this sample relative to the PRG sample studied in \citealt{Ramos11}.

Since no observations of photometric standard stars were taken during the OBEY survey observations, 
we self-calibrated the images using aperture photometry measurements of the sample of elliptical galaxies
from \citet{Prugniel98}, as in \citet{Tal09}.

\subsection{The Extended Groth Strip sample} 
\label{egs}

In order to develop a control sample for the PRGs in the 2Jy sample at redshifts $0.2\leq z<0.7$
we have used the {\it Rainbow Cosmological Surveys database}\footnote{https://rainbowx.fis.ucm.es/Rainbow$_{-}$Database},    
which is a compilation of photometric and spectroscopic data, jointly with value-added products such as photometric 
redshifts, stellar masses, star formation rates, and synthetic rest-frame magnitudes, for several deep cosmological fields 
\citep{Perez08,Barro09,Barro11}.
Specifically, we have selected our control sample in the EGS ($\alpha$ = 14$^{h}~17^{m}$, 
$\delta$ = +52$\degr~30^{\prime}$), which  enlarges the HST Groth-Westphal strip \citep{Groth94} 
up to 2$\degr\times15^{\prime}$ and has the advantage of
being a low extinction area. We chose the EGS because it is a survey that covers sufficient area, and consequently 
enough galaxies, to extract statistically meaningful results, and because of the vast amount of public data 
available, including photometric redshifts, absolute magnitudes, colours, and deep imaging in the optical \citep{Davis07}. 
Indeed, here we use broadband images in the R$_c$ filter obtained with the 
Subaru Telescope, which are similar in pixel size and depth to the GMOS-S images of the PRGs employed in \citealt{Ramos11}.

Thus, we selected all the galaxies in the EGS with the same redshift and absolute magnitude ranges as the PRGs 
at $z\geq0.2$ in \citealt{Ramos11} ($0.2\leq z<0.7$ and $-22.2\leq M_B \leq -20.6$ mag respectively). 
The limiting values of this M$_B$ range were defined by considering NLRGs and WLRGs in \citealt{Ramos11}, 
since the quasars and BLRGs are likely to be contaminated by a large contribution 
from AGN emission. From this first selection we discard the sources in the EGS detected in X-rays
(i.e. possible AGN) and foreground stars. The stars were automatically identified based on a combination of several criteria 
including their morphology (stellarity index) and their optical/NIR colours (see \citealt{Perez08} and \citealt{Barro11} for details on 
the star-galaxy separation criteria). 
In order to identify elliptical galaxies, we imposed a colour selection criterion: initially we selected
all the sources with rest-frame colours (M$_u$-M$_g$) $>$ 1.5, typical of galaxies located in the red sequence in the colour-magnitude diagram 
\citep{Blanton06}.

We choose this initial colour selection rather than morphologically selecting elliptical galaxies from the outset in order to avoid possible 
biases. The goal of this paper is to compare the morphologies of quiescent ellipticals with those of PRGs; by morphologically 
selecting elliptical galaxies (either by eye or automatically) we could be discarding highly disturbed sources, leading to 
a underestimation of interacting systems in the control sample. After applying the colour selection, 
we made a first visual classification of the sources into three groups: elliptical galaxies (E), 
possible disks (PD), and disks (D). According to \citet{Bundy10}, the red sequence is populated not only by elliptical and S0 galaxies, 
but also by early-type spirals. 
We then discarded all the galaxies that appeared as clear disks and kept the elliptical galaxies and possible disks in the sample.
The latter might include disturbed ellipticals that look more disk-like, or S0/early-type spirals. 
After considering all these criteria, we have a control sample of 107 red early-type (ET) galaxies in the EGS 
(see Table \ref{data2}). 

\begin{figure}
\centering
\includegraphics[width=8.5cm]{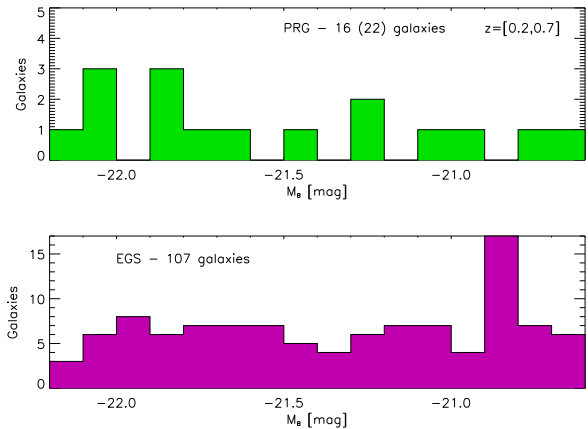}
\caption{Comparison between the B-band absolute magnitudes of the PRGs in the 2Jy sample at 
$0.2\leq z<0.7$ (top panel) and those of the red galaxies in the EGS (bottom panel).}
\label{egs1}
\end{figure}

\begin{figure}
\centering
\includegraphics[width=8.5cm]{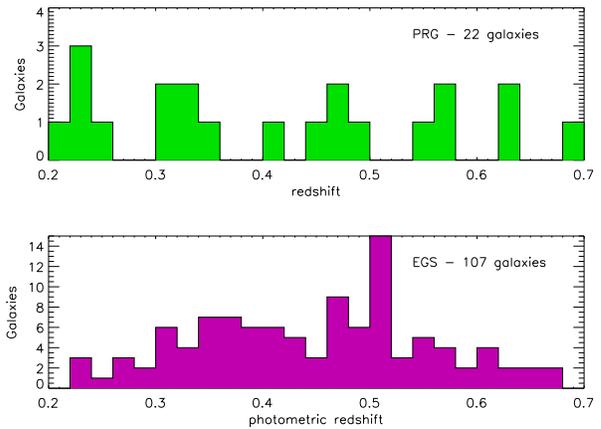}
\caption{Comparison between the spectroscopic redshifts of the PRGs in the 2Jy sample at 
and the photometric redshifts of the red galaxies in the EGS (bottom panel).}
\label{egs2}
\end{figure}

In Figures \ref{egs1} and \ref{egs2} we show the comparison in absolute magnitude and redshift  
between the PRGs and the EGS control sample. Note that in Figure \ref{egs1} we do not include
the 6 BLRGs and quasars with M$_B<-22.2$ mag. The EGS redshifts are photometric
with an average quality of $\Delta$z/(1+z) = 0.03. The absolute magnitudes were estimated 
by convolving the best fitting galaxy templates, used to calculate the photometric redshifts, 
with the appropriate filter transmission (see \citealt{Perez08,Barro09,Barro11} for specific details).
The significance level of the KS statistic from the comparison between the two M$_B$ distributions 
shown in Figure \ref{egs1} is 0.18, indicating that there is no significant difference between 
the two samples in terms of the distribution of absolute magnitude. The same is valid for the 
comparison between the redshift distributions (Figure \ref{egs2}), for which the value
of the KS probability is 0.15.

\begin{figure*}
\centering
\subfigure[]{\includegraphics[width=5cm]{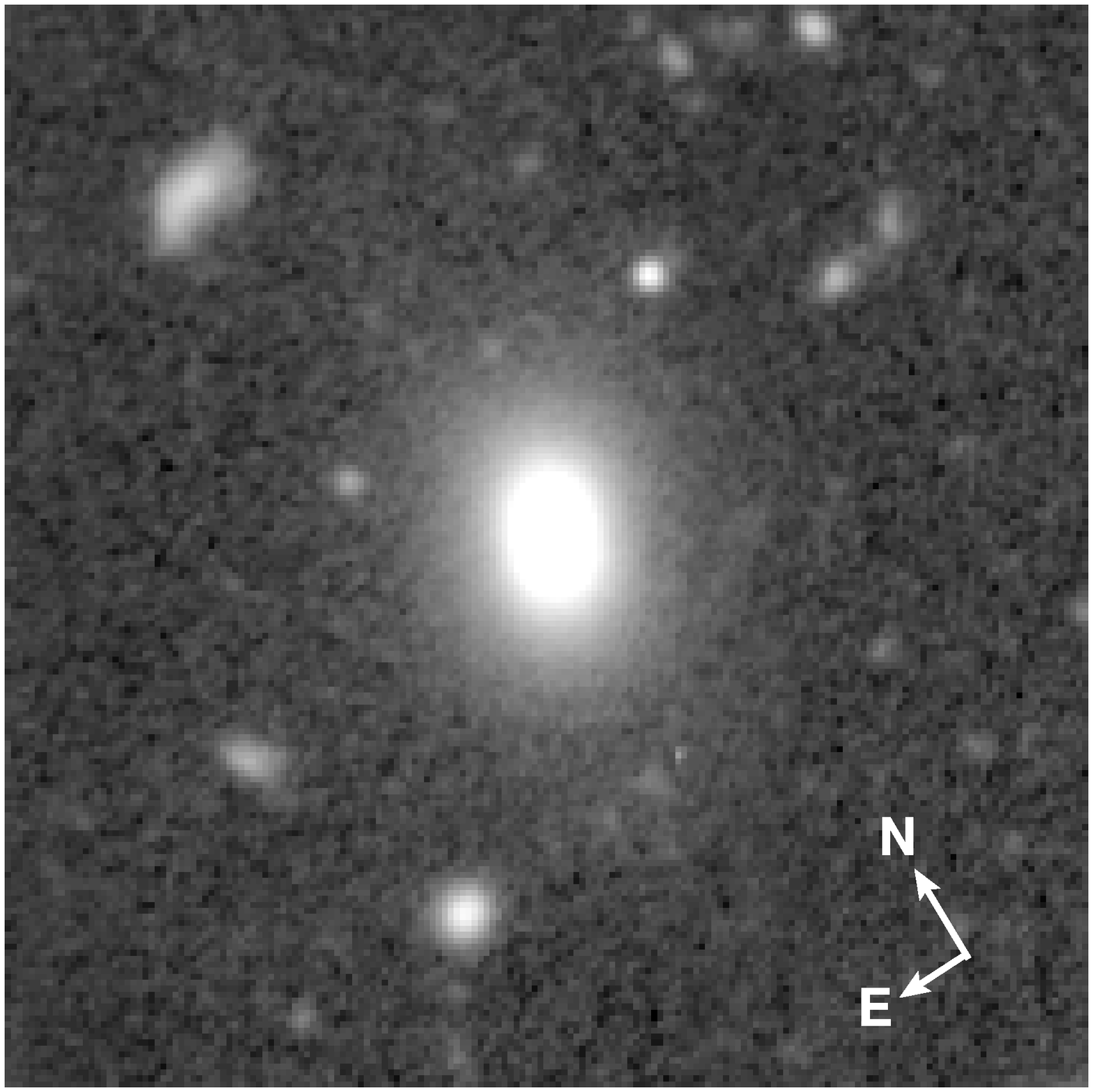}
\label{irac105193}}
\subfigure[]{\includegraphics[width=5cm]{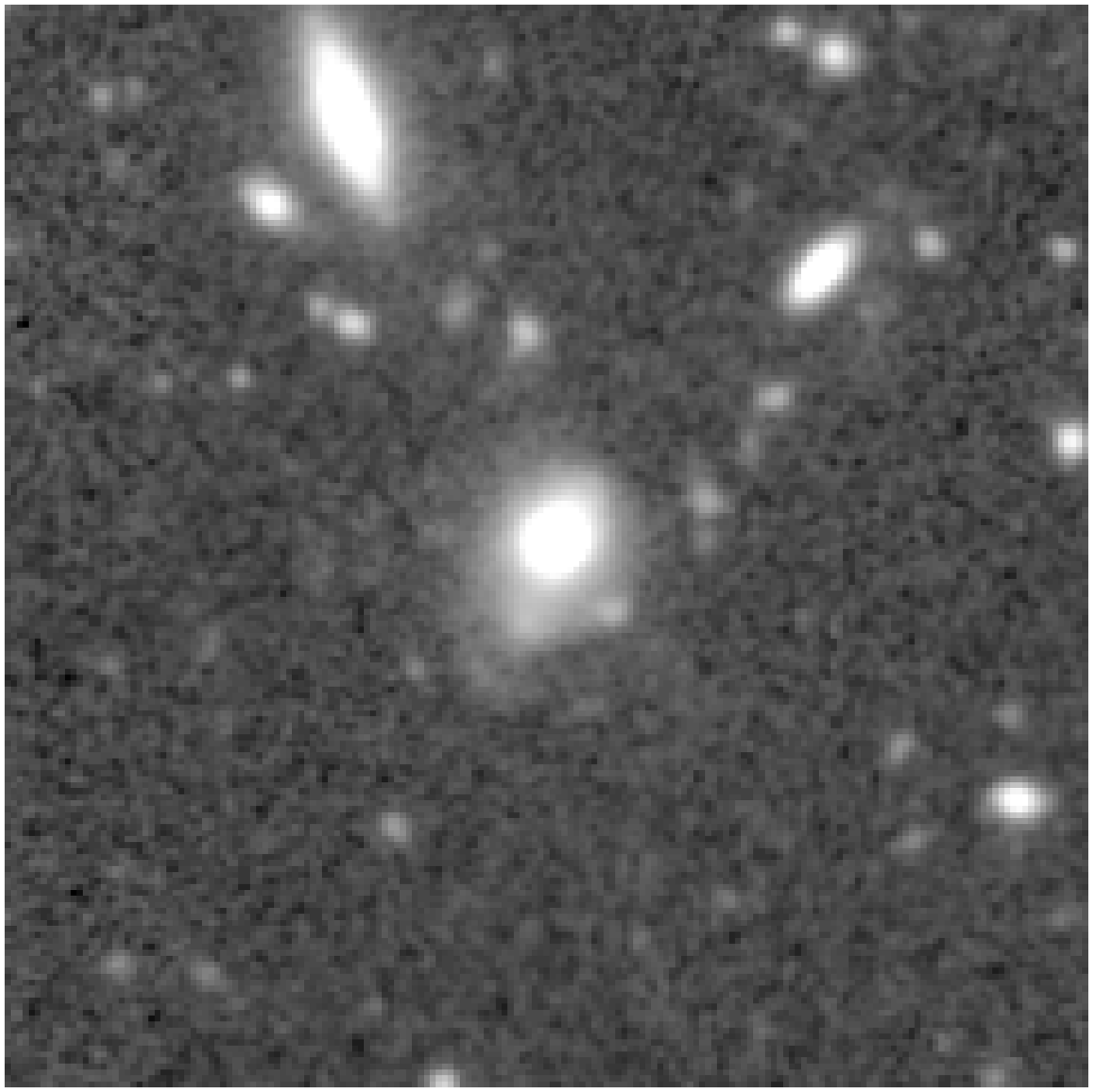}
\label{irac145098}}
\subfigure[]{\includegraphics[width=5cm]{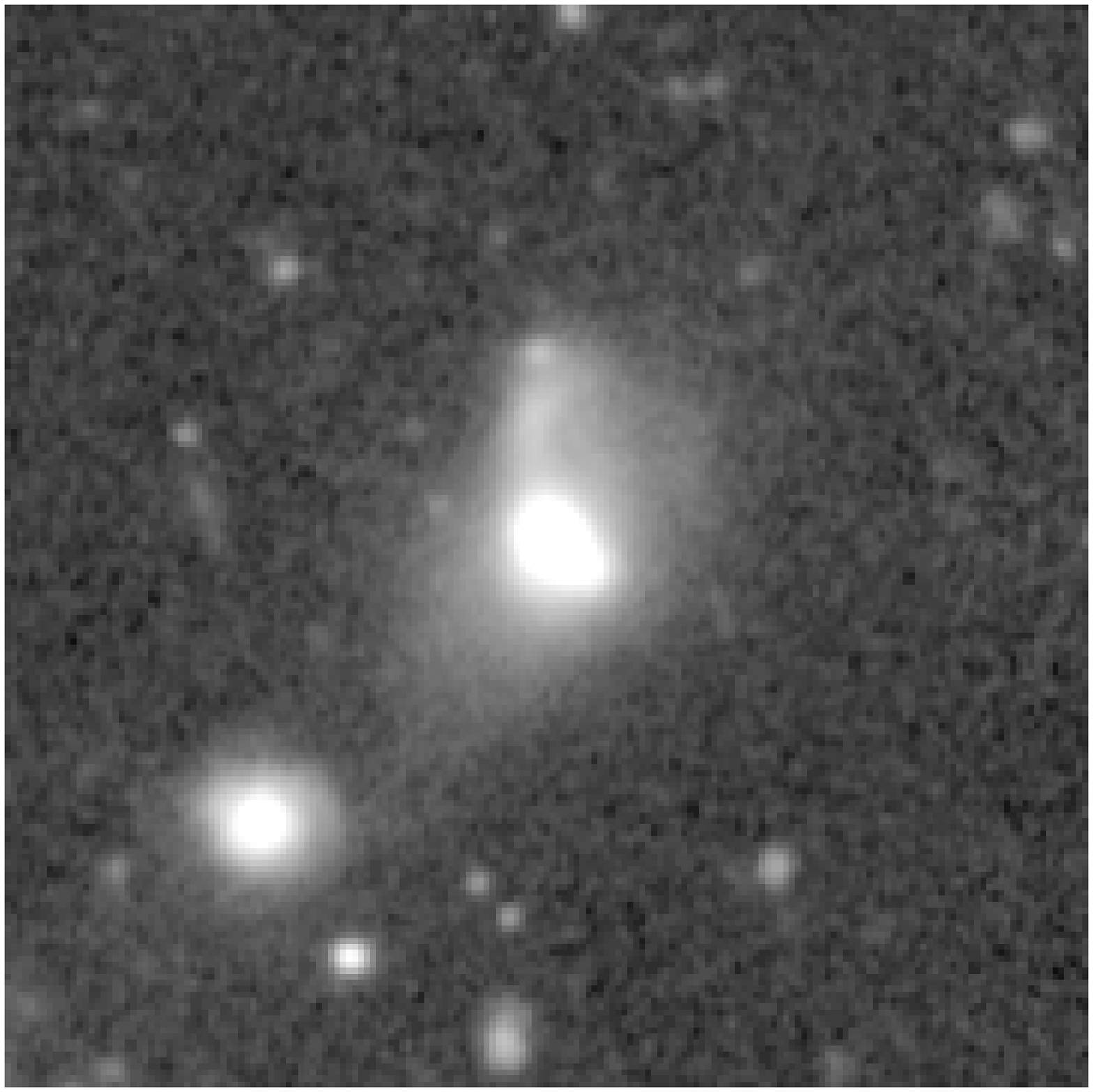}
\label{irac111427}}
\subfigure[]{\includegraphics[width=5cm]{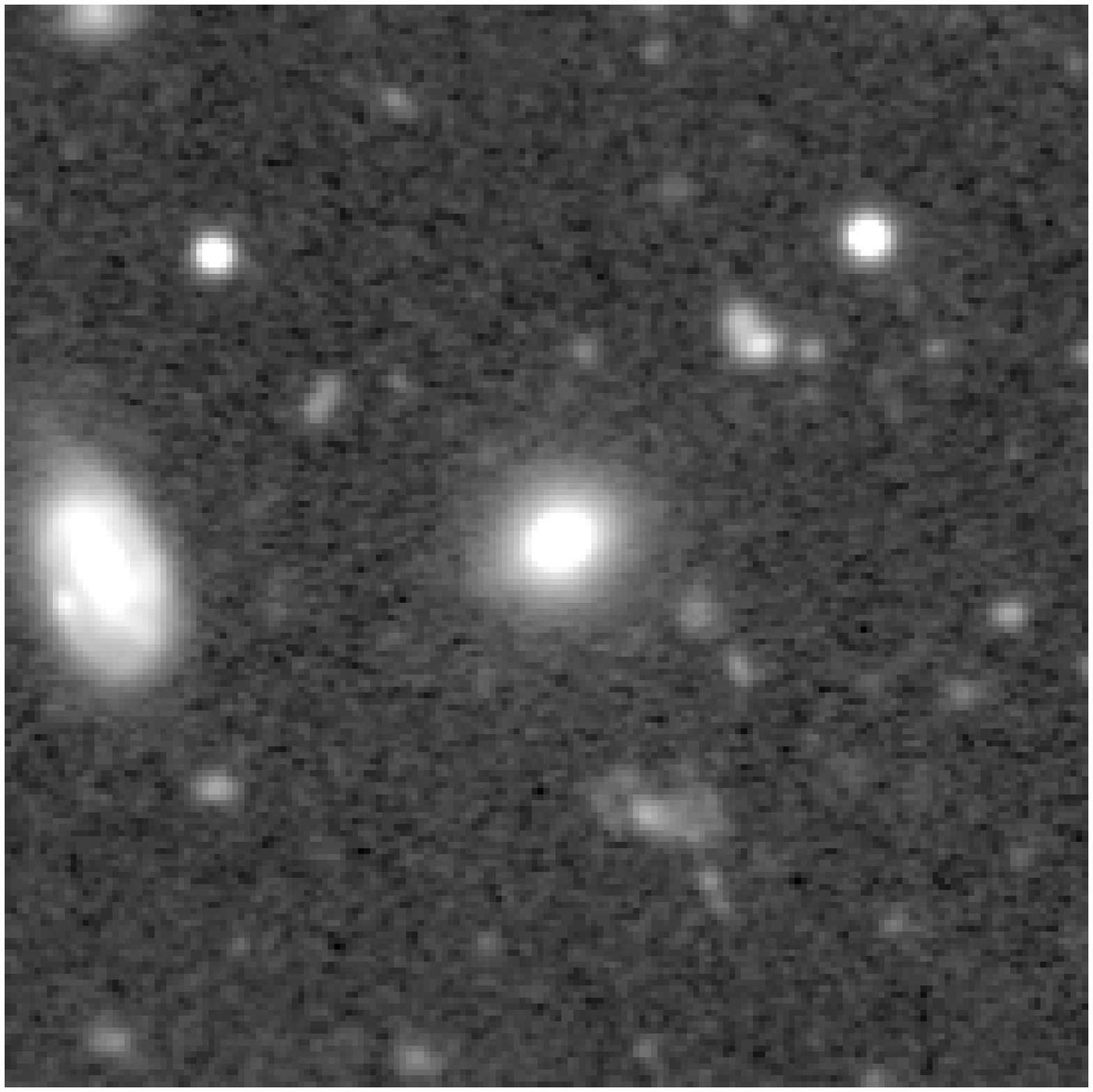}
\label{irac161724}}
\subfigure[]{\includegraphics[width=5cm]{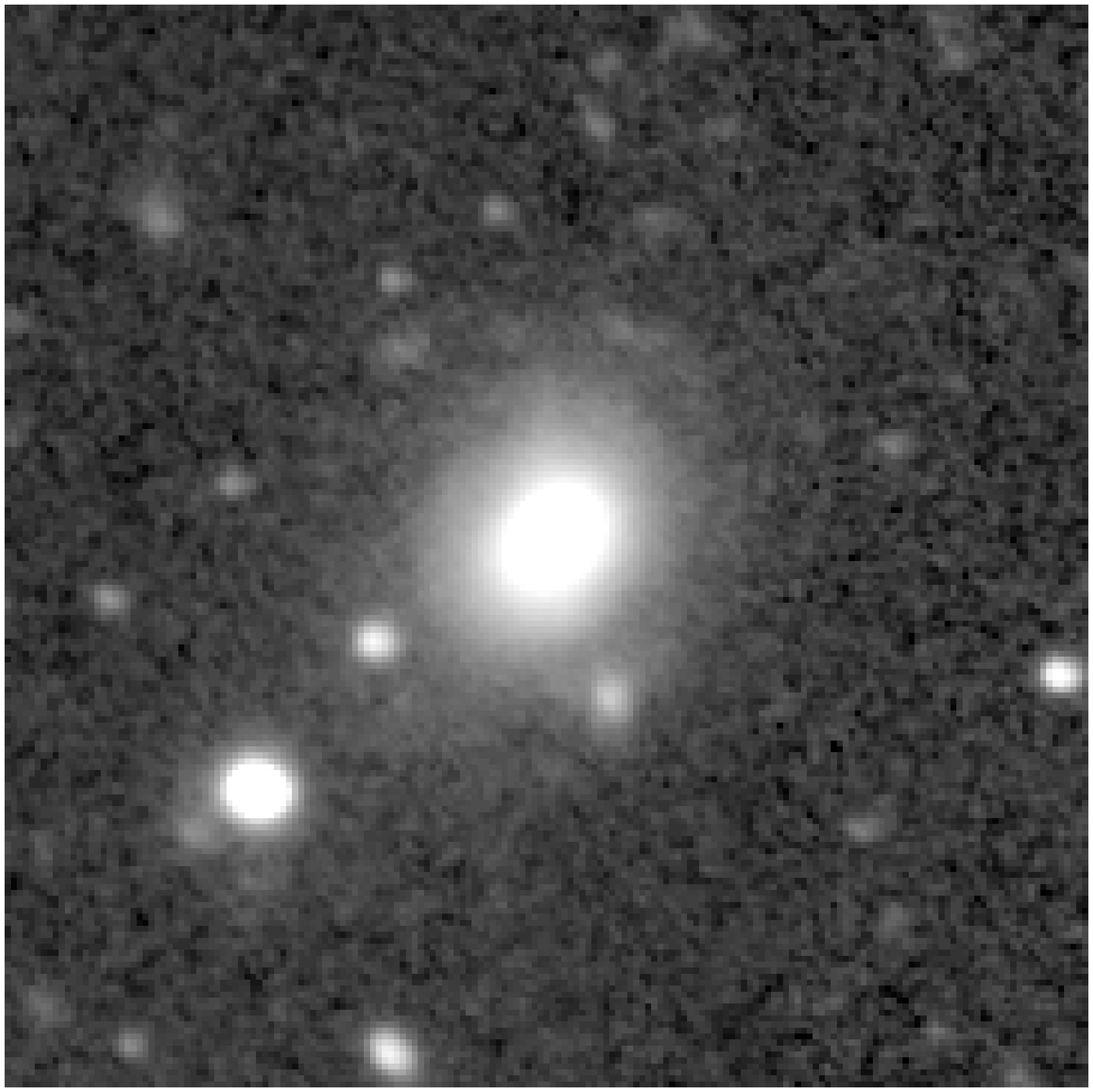}
\label{irac204944}}
\subfigure[]{\includegraphics[width=5cm]{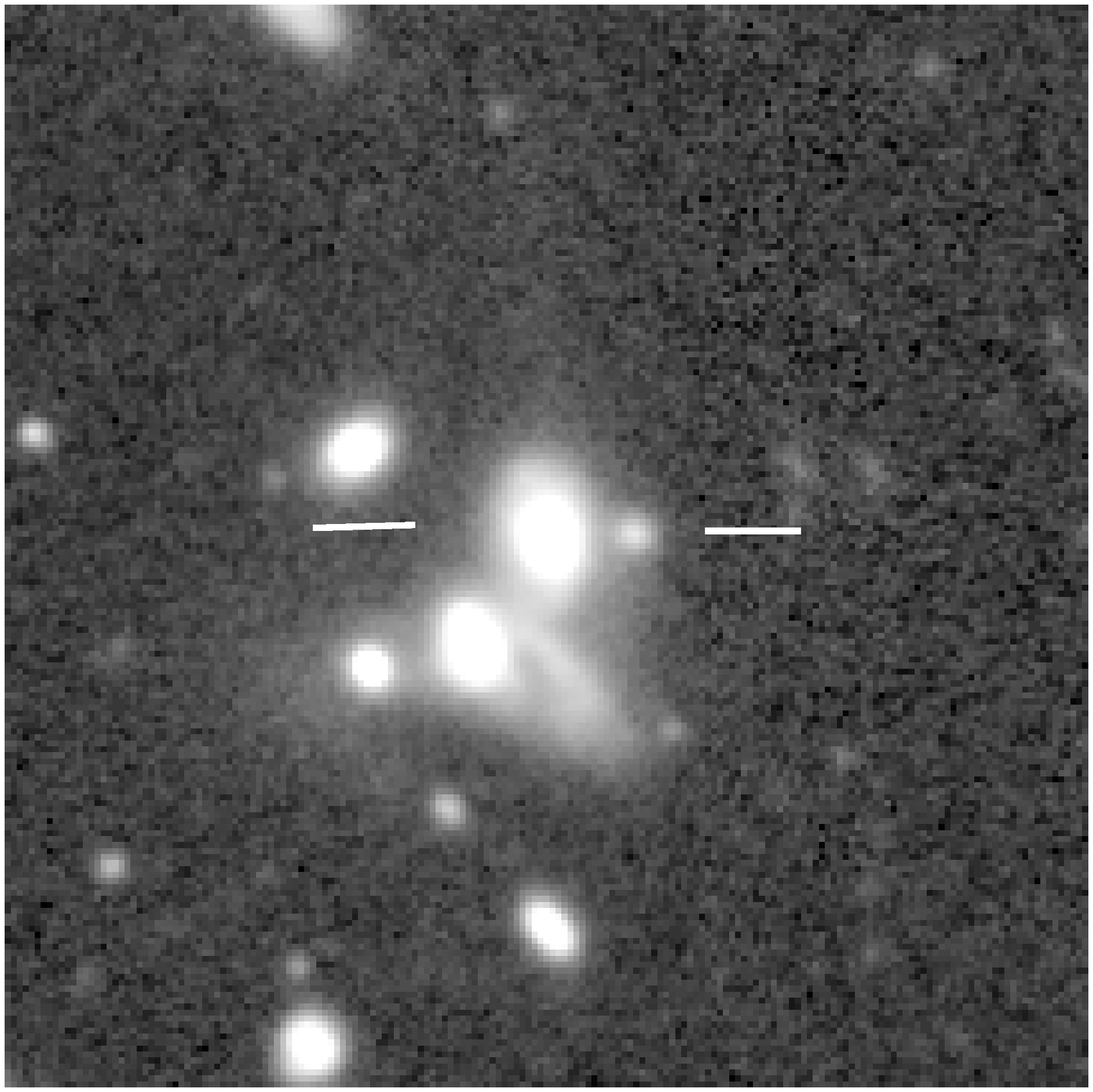}
\label{irac124509}}
\caption{Examples of early-type galaxies in the EGS sample. The level of disturbance
in their morphologies increases from left to right. Images size is 36\arcsec~side. 
IDs and classification from Table \ref{data2} follow: 
(a) irac105193 - undisturbed; (b) irac145098 - A,T; (c) irac111427 - 2N,T,2I; 
(d) irac161724 - undisturbed; (e) irac204944 - T,S; (f) irac124509 - B,2T,F.}
\label{egs_images}
\end{figure*}

The EGS was imaged  with the Subaru Prime Focus Camera (Suprime-Cam; \citealt{Miyazaki02}), 
which is a mosaic of ten 2048$\times$4096 CCDs, located at the prime focus of Subaru Telescope.
Details on the observations of the EGS can be found in \citep{Zhao09}. 
It covers a 34$\times$27 arcmin$^2$ FOV with a pixel scale of 0.202\arcsec. 
The Suprime-Cam data consist of four Rc-band images of 1200 s exposure time that cover the 
entire field to a 5$\sigma$ limiting AB magnitude of $\sim$26.5 \citep{Park08}. In this work
we have measured a surface brightness of $\mu_V$ = 26.3 mag~arcsec$^{-2}$ for the faintest 
feature detected, as shown in Section \ref{mu}.
The seeing measured for the 4 images ranges from 0.60\arcsec to 0.72\arcsec. 
Thus, the data are comparable in depth and resolution to the GMOS-S images employed in the study 
of the morphologies of the PRGs. In Figure \ref{egs_images} we present six examples of Suprime-Cam 
images of galaxies in the EGS, showing different levels of disturbance in their morphologies.

\begin{table*}
\centering
\begin{tabular}{ccccccccc}
\hline
\hline
IRAC ID  &  R.A.        & Dec       & z$_{phot}$& M$_B$    & (M$u$-M$g$) &Type & Morphology  & Group  \\
\hline
004162   &  14:21:29.6  & 52:58:35  &   0.48    &  -20.65  &  1.75  & E   & ...         &    5      \\
006612   &  14:17:29.5	& 52:44:33  &   0.31	&  -21.56  &  1.71  & E   & B,F,[D],[T] &    1,3    \\
006613   &  14:17:29.3	& 52:44:29  &   0.30	&  -20.81  &  1.70  & E   & B		&    1      \\
056690   &  14:15:14.4	& 52:12:52  &   0.50	&  -21.01  &  1.70  & E   & [A],[B]	&    5      \\
060191   &  14:23:43.2	& 53:35:33  &   0.57	&  -21.82  &  1.72  & E   & F		&    2      \\
060958   &  14:23:25.2	& 53:37:59  &   0.40	&  -20.84  &  1.84  & PD  & T,[A],[B]	&    1,2    \\
061249   &  14:23:34.8	& 53:35:28  &   0.65	&  -21.95  &  1.68  & PD  & [T] 	&    5      \\
066105   &  14:23:25.2	& 53:26:14  &   0.51	&  -21.46  &  1.64  & E   & [A] 	&    5      \\
067417   &  14:23:01.4	& 53:28:33  &   0.39	&  -21.25  &  1.60  & E   & ... 	&    5      \\
072533   &  14:22:26.9	& 53:25:19  &   0.33	&  -20.67  &  1.64  & E   & S		&    2      \\
073519   &  14:22:21.1	& 53:24:32  &   0.49	&  -21.20  &  1.55  & E   & [A] 	&    5      \\
074777   &  14:22:24.5	& 53:21:09  &   0.42	&  -21.91  &  1.61  & E   & [S] 	&    5      \\
074924   &  14:22:24.5	& 53:20:53  &   0.41	&  -21.35  &  1.67  & E   & ... 	&    5      \\
077695   &  14:22:18.2	& 53:16:39  &   0.35	&  -20.92  &  1.69  & PD  & T		&    2      \\
079968   &  14:22:02.4	& 53:15:17  &   0.60	&  -21.62  &  1.82  & E   & F		&    2      \\
082325   &  14:21:22.8	& 53:18:39  &   0.55	&  -22.05  &  1.82  & E   & [F] 	&    5      \\
083714   &  14:21:24.5	& 53:15:34  &   0.50	&  -22.01  &  1.52  & E   & F		&    2      \\
088031   &  14:21:17.5	& 53:09:09  &   0.50	&  -21.04  &  1.68  & E   & F		&    2      \\
090430   &  14:21:09.1	& 53:07:43  &   0.38	&  -21.83  &  1.75  & E   & A,F,[B]	&    2      \\
092065   &  14:21:10.3	& 53:05:40  &   0.55	&  -21.44  &  1.80  & E   & B		&    1      \\
092765   &  14:20:41.0	& 53:11:01  &   0.35	&  -21.60  &  1.77  & PD  & [A],[T]	&    5      \\
093764   &  14:20:34.3	& 53:11:23  &   0.39	&  -21.50  &  1.52  & E   & [S] 	&    5      \\
094231   &  14:21:00.0	& 53:05:36  &   0.41	&  -21.98  &  1.81  & E   & 2F,[T]	&    2      \\
094966   &  14:21:21.1	& 53:00:27  &   0.46	&  -21.65  &  1.76  & E   & 2T  	&    2      \\
095727   &  14:20:48.7	& 53:06:24  &   0.38	&  -21.59  &  1.76  & E   & F,S 	&    2      \\
099954   &  14:20:14.4	& 53:09:06  &   0.27	&  -20.83  &  1.71  & E   & [T] 	&    5      \\
102757   &  14:19:57.6	& 53:09:40  &   0.22	&  -20.81  &  1.71  & E   & 2S  	&    2      \\
102982   &  14:20:56.6	& 52:57:14  &   0.60	&  -21.80  &  1.67  & E   & F		&    2      \\
103198   &  14:20:43.2	& 52:59:46  &   0.38	&  -21.45  &  1.76  & E   & 2N,F,S	&    2,3    \\
104038   &  14:20:29.3	& 53:01:46  &   0.46	&  -20.72  &  1.75  & E   & B		&    1      \\
104729   &  14:20:43.0	& 52:58:14  &   0.63	&  -21.76  &  1.61  & PD  & A		&    2      \\
105193   &  14:20:07.0	& 53:05:07  &   0.23	&  -21.08  &  1.90  & E   & [S] 	&    5      \\
106324   &  14:19:48.0	& 53:07:49  &   0.26	&  -20.85  &  1.60  & E   & [T] 	&    5      \\
106984   &  14:20:47.0  & 52:54:57  &   0.45	&  -20.98  &  1.72  & PD  & A,[I]	&    2      \\
111427   &  14:20:23.5  & 52:55:02  &   0.32	&  -20.81  &  1.66  & E   & 2N,T,2I	&    1,3    \\
112580   &  14:20:01.2  & 52:58:23  &   0.51	&  -21.62  &  1.70  & E   & [B] 	&    5      \\
113088   &  14:20:00.7  & 52:57:57  &   0.48	&  -21.22  &  1.76  & E   & [B] 	&    5      \\
113577   &  14:19:56.4  & 52:58:21  &   0.67	&  -22.05  &  1.56  & PD  & [A] 	&    5      \\
114966   &  14:20:25.1	& 52:50:53  &   0.61    &  -22.18  &  1.69  & PD  & 2T,S        &    2      \\ 
115327   &  14:19:37.2	& 53:00:20  &   0.35	&  -20.96  &  1.76  & E   & 2F,[T]	&    2      \\
115594   &  14:20:27.8	& 52:49:36  &   0.31	&  -20.81  &  1.78  & E   & 2N,T	&    2,3    \\
118942   &  14:20:21.8	& 52:47:15  &   0.37	&  -21.04  &  1.63  & E   & ... 	&    5      \\
119696   &  14:20:18.2	& 52:47:12  &   0.50	&  -21.77  &  1.68  & E   & B,F 	&    1,2    \\
122098   &  14:19:26.6	& 52:55:17  &   0.22	&  -21.34  &  1.65  & PD  & ... 	&    5      \\
124509   &  14:19:29.8	& 52:51:59  &   0.34	&  -20.75  &  1.84  & PD  & B,2T,F	&    1,2    \\
125663   &  14:19:34.8	& 52:49:47  &   0.53	&  -21.53  &  1.59  & E   & [F] 	&    5      \\
126918   &  14:18:57.1	& 52:56:12  &   0.49	&  -21.22  &  1.77  & E   & F,[B]	&    1,2    \\
127241   &  14:20:00.0	& 52:42:54  &   0.59	&  -21.75  &  1.67  & PD  & ... 	&    5      \\
127457   &  14:18:47.0	& 52:57:40  &   0.50	&  -22.08  &  1.63  & E   & 2N,A	&    2,3    \\
128074   &  14:19:09.6	& 52:52:25  &   0.34	&  -20.73  &  1.65  & E   & B,[F]	&    1      \\
128416   &  14:19:58.3	& 52:42:01  &   0.58	&  -21.51  &  1.68  & PD  & ... 	&    5      \\
132682   &  14:18:39.8	& 52:53:49  &   0.33	&  -20.83  &  1.51  & E   & ... 	&    5      \\
135859   &  14:18:49.0	& 52:48:38  &   0.40	&  -20.88  &  1.80  & E   & [I] 	&    5      \\
\hline		     			      		 
\end{tabular}						 
\end{table*}						 

\begin{table*}
\centering
\begin{tabular}{ccccccccc}
\hline
\hline
IRAC ID  &  R.A.        & Dec       & z$_{phot}$& M$_B$    & (M$u$-M$g$) &Type & Morphology  & Group  \\
\hline
138794   &  14:19:17.5	& 52:39:40  &   0.50	&  -21.23  &  1.70  & E   & [T] 	&    5      \\
139190   &  14:19:00.2	& 52:42:49  &   0.44	&  -20.66  &  1.66  & E   & ... 	&    5      \\
140456   &  14:19:08.2	& 52:39:53  &   0.30	&  -21.13  &  1.78  & PD  & 2T  	&    2      \\
140758   &  14:19:15.4	& 52:38:05  &   0.43	&  -21.31  &  1.68  & E   & S		&    2      \\
141714   &  14:18:47.3  & 52:42:51  &   0.44    &  -20.65  &  1.76  & PD  & [B],[S]     &    5      \\
143149   &  14:18:07.9	& 52:49:24  &   0.37	&  -21.81  &  1.77  & E   & T		&    2      \\
143536   &  14:18:33.1	& 52:43:52  &   0.50	&  -21.20  &  1.81  & PD  & [T] 	&    5      \\
145098   &  14:18:10.6	& 52:46:50  &   0.32	&  -20.84  &  1.70  & E   & A,T 	&    2      \\
145434   &  14:18:53.3	& 52:37:43  &   0.48	&  -21.68  &  1.68  & PD  & 4T  	&    2      \\
146298   &  14:18:35.0	& 52:40:34  &   0.59	&  -21.57  &  1.76  & E   & [A] 	&    5      \\
152722   &  14:17:41.3	& 52:44:45  &   0.49	&  -20.77  &  1.55  & PD  & [F] 	&    5      \\
156161   &  14:18:36.5	& 52:29:35  &   0.30	&  -20.71  &  1.79  & E   & T		&    2      \\
157751   &  14:18:24.5	& 52:30:24  &   0.47	&  -21.22  &  1.81  & E   & ... 	&    5      \\
157878   &  14:17:30.2	& 52:41:20  &   0.46	&  -20.85  &  1.70  & E   & F		&    2      \\
159123   &  14:18:15.8	& 52:30:37  &   0.56	&  -22.03  &  1.70  & PD  & T		&    2      \\
159936   &  14:17:28.3	& 52:39:26  &   0.41	&  -21.22  &  1.55  & E   & 2N  	&    3      \\
160442   &  14:17:33.8	& 52:37:46  &   0.47	&  -21.90  &  1.61  & E   & B,A 	&    1,2    \\
160500   &  14:17:33.1	& 52:37:53  &   0.34	&  -20.81  &  1.78  & PD  & B,2T	&    1,2    \\
161724   &  14:17:25.4	& 52:38:08  &   0.34	&  -20.71  &  1.95  & PD  & [F] 	&    5      \\
165265   &  14:17:11.0	& 52:37:29  &   0.67	&  -22.04  &  1.55  & E   & B,T 	&    1,2    \\
166730   &  14:17:53.5	& 52:27:22  &   0.36	&  -21.01  &  1.55  & E   & S,T 	&    2      \\
169386   &  14:16:58.6	& 52:35:49  &   0.47	&  -20.68  &  1.55  & PD  & ... 	&    5      \\
172474   &  14:17:19.9	& 52:28:24  &   0.51	&  -20.90  &  1.82  & E   & T,B,F	&    1,2    \\
173901   &  14:17:32.2	& 52:24:15  &   0.32	&  -21.39  &  1.76  & E   & ... 	&    5      \\
175347   &  14:17:08.9	& 52:27:09  &   0.60	&  -21.92  &  1.60  & E   & S,[B]	&    2      \\
175590   &  14:17:15.4	& 52:25:33  &   0.56	&  -21.68  &  1.71  & E   & [A] 	&    5      \\
177990   &  14:16:41.3	& 52:29:02  &   0.25	&  -21.05  &  1.75  & PD  & F,[2N]	&    2      \\
178118   &  14:17:20.2	& 52:20:51  &   0.46	&  -21.74  &  1.75  & E   & ... 	&    5      \\
178724   &  14:17:15.1	& 52:20:51  &   0.52	&  -21.81  &  1.67  & E   & A		&    2      \\
178868   &  14:16:57.5  & 52:24:09  &   0.37    &  -22.14  &  1.69  & PD  & F           &    2      \\ 
180420   &  14:16:54.0	& 52:21:50  &   0.54	&  -21.91  &  1.62  & E   & 2N,2T,[B]	&    2,3    \\
181402   &  14:16:38.2	& 52:23:08  &   0.38	&  -20.87  &  1.76  & E   & [I],[A]	&    5      \\
181444   &  14:16:47.3	& 52:21:11  &   0.31	&  -21.63  &  1.71  & E   & 2S,[I]	&    2      \\
181736   &  14:16:27.4	& 52:24:39  &   0.46	&  -20.93  &  1.79  & PD  & ... 	&    5      \\
181914   &  14:16:57.6	& 52:18:10  &   0.36	&  -20.72  &  1.66  & PD  & ... 	&    5      \\
182762   &  14:16:52.8	& 52:17:28  &   0.43	&  -21.41  &  1.71  & PD  & [F] 	&    5      \\
183081   &  14:16:49.0	& 52:17:38  &   0.36	&  -21.86  &  1.71  & E   & F,[T]	&    2      \\
183836   &  14:16:43.2	& 52:17:21  &   0.44	&  -21.02  &  1.76  & E   & [S] 	&    5      \\
184041   &  14:16:40.0  & 52:17:35  &   0.53	&  -22.13  &  1.75  & E   & F,S         &    2      \\
184315   &  14:16:16.8	& 52:21:46  &   0.50	&  -21.61  &  1.64  & PD  & 2N  	&    3      \\
186058   &  14:16:08.9	& 52:19:59  &   0.54	&  -21.77  &  1.80  & PD  & [A] 	&    5      \\
189727   &  14:16:15.1  & 52:11:21  &   0.64	&  -21.96  &  1.80  & PD  & ... 	&    5      \\
190795   &  14:16:10.3  & 52:10:12  &   0.51	&  -21.19  &  1.70  & PD  & T,S 	&    2      \\
193464   &  14:15:36.5	& 52:11:41  &   0.42	&  -20.69  &  1.66  & E   & 2N,F	&    2,3    \\
193507   &  14:16:03.1	& 52:06:11  &   0.47	&  -21.91  &  1.71  & E   & 2N,[B]	&    3      \\
193737   &  14:15:54.5	& 52:07:30  &   0.50	&  -20.88  &  1.55  & E   & ... 	&    5      \\
193974   &  14:15:31.4	& 52:11:46  &   0.40	&  -21.20  &  1.73  & E   & [S] 	&    5      \\
194092   &  14:15:29.0	& 52:12:00  &   0.51	&  -20.88  &  1.66  & E   & [T] 	&    5      \\
196827   &  14:15:41.3	& 52:03:43  &   0.37	&  -20.86  &  1.76  & E   & T		&    2      \\
198295   &  14:14:58.6	& 52:09:25  &   0.54	&  -21.91  &  1.72  & E   & [S] 	&    5      \\
199503   &  14:14:56.2	& 52:07:26  &   0.50	&  -21.20  &  1.70  & PD  & T		&    2      \\
202111   &  14:14:41.3	& 52:04:54  &   0.27	&  -21.16  &  1.70  & E   & [S] 	&    5      \\
204161   &  14:14:57.8	& 51:57:54  &   0.62	&  -21.78  &  1.88  & E   & A,[B]	&    2      \\
204944   &  14:14:41.3	& 51:59:40  &   0.28	&  -21.55  &  1.66  & PD  & T,S 	&    2      \\
\hline		     			      		 
\end{tabular}						 
\caption{Full classification of the EGS sample ordered by IRAC ID (Rainbow database identifier). 
Columns 2, 3, and 4 list R.A., declination and photometric redshift. Columns 5 and 6 correspond to the 
B-band absolute magnitudes and (M$u$-M$g$) rest-frame colors from the Rainbow database. Column 7 indicates whether 
a galaxy has been visually classified as an elliptical or as a possible disk. Columns 8 and 9 list the morphological 
classification and group as in Table \ref{data}. Features with uncertain identification (within brackets) have not 
been considered in the statistics discussed in this work.}
\label{data2}
\end{table*}

\section{Results}

\subsection{Optical Morphologies}
\label{morphology}

The main aim of this work is to perform a morphological classification of the galaxies in the 
OBEY and EGS samples in the same manner as for the PRGs in the 2Jy sample studied in 
\citealt{Ramos11} and then to compare the results. This comparison will allow us to determine 
whether or not galaxy interactions are more common in powerful AGN than in quiescent galaxies and,
consequently, to establish how important these interactions are in the triggering of nuclear
activity.

\subsubsection{Morphological Features}
\label{morphological_features}

The morphological classification of the galaxies was done blind, with no information about any previous 
work on the sources, by CRA visually inspecting the 55 OBEY survey and the 107 EGS images.  
In addition, PB and CT also examined the EGS galaxy morphologies and 
there was agreement among the classifiers for the majority of the galaxies. Any possible
conflicts were resolved by re-examining the images. 
The classification of the various features detected in the two control samples  
is based on that first used by \citet{Heckman86} and is exactly the same as employed in \citealt{Ramos11}.
Note that the classification of the PRGs was done using the fully-reduced GMOS-S original images, 
before any image enhancement technique were applied to them.
The following morphological features are considered.

\begin{itemize}

\item A $tail$ (T) corresponds to a narrow curvilinear feature with roughly radial orientation. 

\item A $fan$ (F) is similar to a tail, but shorter and broader.

\item A $bridge$ (B) is a feature that links the radio galaxy with a companion. 

\item A $shell$ (S) is a curving filamentary structure with a roughly tangential orientation 
to the main body of the galaxy.

\item $Dust$ (D) includes both nuclear dusty features and large scale dust lanes. 

\item $Amorphous$ $haloes$ (A) are irregular galaxy hosts or inner features that cannot be 
clearly distinguished from the main body of the galaxy (e.g. knotty haloes).

\item By $irregular$ (I) we refer to any feature, generally elongated, that cannot be classified as any of the previous.
 
\item $Double~nuclei$ (2N) are those composed of two bright peaks inside 10 kpc, following the 
definition employed by \citet{Smith89}, based on statistical studies of cluster galaxies \citep{Hoessel80}\footnote{\citet{Hoessel80} 
claimed that typical cluster members are expected to experience a close encounter or merger within this radius every 10$^9$ years. } 
and N-body simulations of interacting binary galaxies \citep{Borne84}. 

\end{itemize}

All of these features, with the possible exception of the dust, are very likely the result 
of galaxy interactions. 
Simulations have shown how spiral-spiral (S-S), elliptical-spiral (E-S) and elliptical-elliptical (E-E)
interactions can produce all of the features that form the basis of our classification \citep{Quinn84,Hernquist92,Cattaneo05,Lotz08,Feldmann08}. For a more detailed description of how the 
different features described above are produced, according to simulations, see Section 5.1.1. in \citealt{Ramos11}.
The classified features for both the OBEY survey and the EGS sample are listed in Tables \ref{data} and \ref{data2}
respectively. Features with uncertain identification (between brackets in Tables \ref{data} and \ref{data2}) have not been considered 
for the statistics. Examples of the EGS galaxy morphologies are shown in Figure \ref{egs_images} and all the sample images
can be individually viewed online in the Rainbow Database\footnote{https://rainbowx.fis.ucm.es/Rainbow$_{-}$navigator$_{-}$public/}. 
For the images of the OBEY survey we refer the reader to \citet{Tal09}.

The projected linear scales of the tidal features reported in Tables \ref{data} and \ref{data2} 
range from less than 10 kpc in the case of galaxies with double nuclei, to $\sim$80-85 kpc in the case
of long tidal tails such as that in NGC 1209 (OBEY survey) and bridges linking galaxies as in the 
case of irac128074 (EGS sample). For the PRGs in the 2Jy sample the longest feature that we measured
was the spectacular bridge in PKS 0349-27, which links the radio galaxy with a distorted conpanion at 
$\sim$83 kpc (\citealt{Ramos11}).

\subsubsection{Quantitative versus visual analysis of tidal disturbance}

In Section \ref{morphological_features} we described how we performed the visual classification of 
the optical morphologies of the PRG, OBEY, and EGS samples. 
In the following, we compare the results of this visual classification 
of the galaxies in the OBEY survey with the quantitative analysis of the degree of tidal disturbance 
carried out by \citet{Tal09} for the same galaxies. The latter authors fitted an elliptical galaxy model 
to the targets and, after masking the foreground 
stars and background galaxies in the sky-subtracted and flat-fielded images, 
they divided the masked frames by the galaxy model. This process 
produces an image of the residuals that they translated into a number, the tidal parameter, defined as:
T=$\overline{|(I_{x,y}/M_{x,y})-1|}$, where I$_{x,y}$ and M$_{x,y}$ are the pixel values at (x,y) of 
the galaxy and model images, respectively. They finally applied a correction for the residual noise 
to the latter values, to derive the corrected tidal parameter (T$_c$; see \citealt{Tal09} for a more 
detailed description of the methodology).

\begin{figure}
\centering
\includegraphics[width=8.7cm]{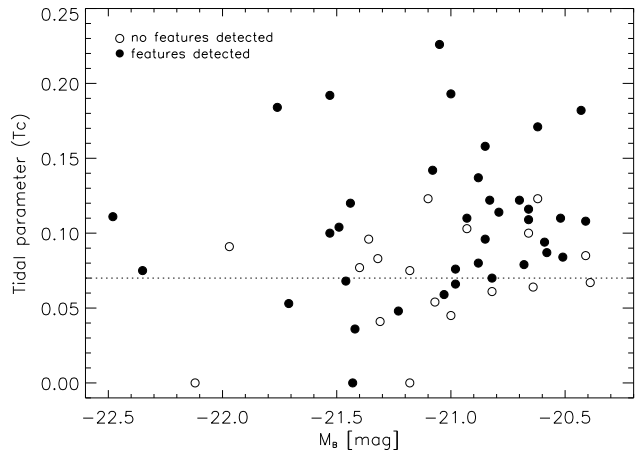}
\caption{Corrected tidal parameter values from \citet{Tal09} for the OBEY survey versus B-band absolute magnitudes.
The dotted horizontal line at T$_c$=0.07 is the boundary between galaxies with and without signs of disturbance from \citet{Tal09}. 
Filled circles correspond to the galaxies visually classified as disturbed in this work, whereas open circles indicate
abcense of morphological features.}
\label{tal}
\end{figure}

Based on the values of T$_c$ determined for the galaxies in the OBEY survey, \citet{Tal09} 
divided them into three groups: 1) galaxies showing clear signs of morphological disturbance (T$_c>0.09$; 
53\% of the sample), 2) galaxies with marginal disturbance (0.07$<$T$_c<0.09$; 20\%), and 3) galaxies
lacking interaction signatures (T$_c<0.07$; 27\%). In Figure \ref{tal} we represent these 
tidal parameters versus the B-band absolute magnitudes (see also Table \ref{data}). First, we note that
there is no clear correlation between the luminosities of the galaxies and the level of disturbance of their
hosts. Second, we have identified with filled circles the galaxies in the OBEY survey for which we 
have found signs of interaction based on our visual classification. From Figure \ref{tal} it is clear
that quantitative and visual classifications do not agree completely. There is complete agreement only 
for the nine galaxies with T$_c>0.13$. In fact, for five galaxies that \citealt{Tal09} classified as 
clearly disturbed (T$_c>0.09$) we do not find any sign of a past interaction. After inspecting the 
images of these five galaxies we conclude that it is likely that the high value of T$_c$ is due to residuals 
from the masking of the foreground stars and background galaxies and/or from isophotal twisting. 
The opposite is also true: we find disturbed morphologies in 7 galaxies with T$_c<0.07$, which are 
either faint relative to their host galaxies or diffuse features that the automatic method misses. 
Overall, however, the total percentages of disturbance agree well between the automatic  
(73\%) and the visual classification (67\%). This vindicates the use of visual rather than quantitative
detection of disturbed morphologies in this paper. 


\subsubsection{Classification}
\label{classification}

Considering the morphological features detected in the OBEY and EGS galaxies (only those with secure 
identifications in Tables \ref{data} and \ref{data2}), the sample can be divided into the following five groups: 

\begin{enumerate}
\renewcommand{\theenumi}{(\arabic{enumi})}

\item {\it Galaxy pair or group in tidal interaction.} Galaxy pairs showing bridges, or co-aligned distorted 
structures. 

\item {\it Galaxies showing tidal features.} Galaxies showing shells, fans, tails, 
amorphous haloes, and irregular features. 

\item {\it Multiple nuclei.} Galaxies with a companion lying inside a 10 kpc radius, according to 
the theoretical definition employed by \citet{Hoessel80} and \citet{Smith89}. 

\item {\it Dust features.} Galaxies presenting dust features as the only sign of disturbance. 

\item {\it Isolated galaxies with no sign of interaction}. 
Objects in which we cannot confidently identify morphological peculiarities.

\end{enumerate}

Note that these categories are not exclusive because some galaxies show more than one of the morphological 
features described above (see Table \ref{data}).
Initially we considered objects in groups 1, 2, 3, and 4 as showing disturbed morphologies consistent with them 
having been involved in a galaxy interaction/merger, whilst galaxies classified in 
the fifth group were classified as undisturbed. Based on this classification, in \citealt{Ramos11} we found that 
85\% of the PRG sample are very likely interacting 
objects or the result of a past merger event. However, $dust$ features by themselves may not necessarily 
be a sign of galaxy interactions. Note that, while small-scale dust is often taken as an observational signature for
recent mergers (e.g., \citealt{vandokkum95}), it may also be associated with cooling flows in central cluster galaxies 
(e.g., \citealt{Fabian94,Hansen95,Edge99,Edge10}).
If we do not consider dust as a sign of morphological disturbance related to mergers and interactions, 
then the percentage of PRGs in the 2Jy sample presenting evidence for interactions is 78\%. 
In Table \ref{groups} we show the percentages of interacting galaxies (those classified in groups 1, 2, or 3) 
for i) the PRGs sample at $z<0.2$ and the OBEY survey, and ii) the PRGs sample at $0.2\leq z<0.7$ and the EGS sample. 

Thus, if we only consider the galaxies classified in groups 1, 2, and 3, 
we find that 62\% of the PRGs at $z<0.2$ show signs of interactions. 
However, it is worth considering the percentage of disturbance after excluding the WLRGs
(shown between parentheses in Table \ref{groups}). 
As explained in the Introduction, it has been proposed that WLRGs are powered by hot gas accretion from their 
X-ray coronae, rather than by the classical AGN cold gas accretion (see also Section 5.2.2. in \citealt{Ramos11}). 
Thus, if we consider SLRGs only, the percentage of PRGs showing signs of interactions increases to 93\%, 
which is higher than the 67\% that we measure for the elliptical galaxies in the OBEY survey. 
On the other hand, 95\% of the PRGs at $0.2\leq z<0.7$ (either including or excluding the only WLRG in this redshift range) 
show signs of interaction, compared with 55\% in the EGS sample (57\% if we consider only ellipticals and not 
the possible disks; column 7 in Table \ref{data2}). Thus, there appears to be more evidence for galaxy interactions
in the PRG sample than in the control samples at all redshifts, although the relatively small size of the PRG 
sample means that the difference is only significant at the $\sim$2$\sigma$ level (for the comparison between the 
$0.2\leq z<0.7$ PRG and EGS samples). However, it is necessary to consider the surface brightnesses of the features 
in order to make a proper comparison with the control sample morphologies, since we know that, for example, the OBEY 
images reach much lower effective surface brightnesses than our GMOS-S images of the PRGs. This comparison provides firmer evidence for
differences between the PRG and quiescent elliptical galaxy samples and it is discussed in Section \ref{mu}.

In terms of the merger scenario, we appear to be observing some 
of the galaxies in the PRG and the two control samples {\it before} the final
coalescence of the nuclei of the interacting system. The galaxies classified in 
groups 1 ({\it galaxy pair or group in tidal interaction}) and 3 ({\it multiple nuclei}) would correspond to
systems observed before the coalescence of the merging nuclei, whereas those in group 2 ({\it galaxies presenting any sign of 
disturbance}) would correspond to more evolved systems (coalescence or post-coalescence). 
The percentage of galaxies in the PRG, OBEY, and EGS samples belonging to each of these groups is shown in 
Table \ref{groups}. 

In \citealt{Ramos11} we showed that, if PRGs are triggered as a consequence of an interaction between galaxies,
it is possible for this to happen any stage of the interaction (before, during, or after the two galaxy nuclei coalesce; 
see also \citealt{Tadhunter11} on basis of young stellar population properties). In fact, we found that
more than one-third of the PRGs in the full 2Jy sample are observed in a pre-coalescence phase (galaxies classified 
in either group 1 or 3). Interestingly, for both the low and intermediate redshift sub-samples, we
find that the proportion of PRGs that are in the pre-coalescence phase is a factor 2--3 higher than that
of the quiescent elliptical galaxies in a similar phase (see Table \ref{groups}). There is also evidence for an increase 
in the proportion of pre-coalescence systems with redshift for both the PRG and quiescent samples (see Table \ref{groups}). 
However, observations of larger samples will be required to put those trends on a more solid statistical footing.

\begin{table*}
\centering
\begin{tabular}{lccccc}
\hline
\hline
Morphology & Group & \multicolumn{1}{c}{PRG Sample $z<0.2$} & \multicolumn{1}{c}{OBEY survey} & \multicolumn{1}{c}{PRG Sample $0.2\leq z<0.7$} & 
\multicolumn{1}{c}{EGS sample} \\
\hline
Signs of interaction		 & 1,2,3&  62\% (93\%) &  67\%  &  95\% (95\%)  & 55\% (57\%)  \\ 		   
\hline
Pre-coalescence                  & 1,3  &  21\% (21\%) &   7\%  &  50\% (50\%)  & 20\% (24\%)  \\
Coalescence or post-coalescence  & 2*   &  42\% (72\%) &  60\%  &  45\% (45\%)  & 35\% (33\%)  \\
No signs of interaction          & 4,5  &  37\%  (7\%) &  33\%  &   5\%  (5\%)  & 45\% (43\%)  \\
\hline		     			      		 
\end{tabular}						 
\caption{Classification of all the galaxies in the PRG, OBEY and EGS samples. Sources
belonging to groups 1 and 3 are considered as pre-coalescence systems, those in group 2 are likely coalescence or post-coalescence 
scenarios, and finally, galaxies classified in groups 4 and 5 do not show signs of interaction. * Those galaxies classified as (2,3) or (1,2) in Tables \ref{data}, \ref{data2}, and Table 1 in \citealt{Ramos11} are considered as pre-coalescence systems here, although they belong to group 2 as well.
Percentages between parentheses correspond to SLRGs in the PRG sample (columns 3 and 5) and to elliptical galaxies only in the EGS sample (column 6).}
\label{groups}
\end{table*}

The lack of dust features in the control sample as compared to the PRG sample is also interesting. For the OBEY
survey we find that only 7\% of the elliptical galaxies show dust, whereas for the PRGs at $z<0.2$ the percentage
increases to 25\% (21\% for the SLRGs). This agrees with the $\sim$30\% of dust features found from 
optical HST images of radio galaxies at $z<0.5$ \citep{deKoff96,deKoff00}.
Dust can either be produced by mass loss from evolving red giant stars \citep{Knapp89,Athey02},
or accreted during galaxy interactions \citep{Goudfrooij94,vandokkum95}. 
If the dust is accreted in mergers/interactions, then we expect to find a higher incidence of dust features in 
the PRGs than in the quiescent ellipticals, mirroring the difference found for signs of disturbance in general
between the two populations (see Table \ref{same}). 
On the other hand, at higher redshifts neither the galaxies in the EGS 
nor in the SLRG sample at $0.2\leq z<0.7$ show dust features. 
This apparent lack of dust at higher redshifts in both the radio and the quiescent galaxies 
is likely to be due to a resolution effect, since dust lanes, with typical scales of $\sim$1-5 kpc, will be poorly 
resolved in the ground-based data employed here and in \citealt{Ramos11} for galaxies at such redshifts. 


We have classified the galaxies in the OBEY and EGS samples in the same manner as the PRGs and   
compared them. However, as we mentioned in Section \ref{obey}, from the comparison between the
M$_B$ histograms of the PRGs at $z<0.2$ and the OBEY sample shown in Figure \ref{tal1}, 
there is only a 4\% chance that the two 
distributions are drawn from the same parent population according to the KS test. In order to 
confirm that this difference does not produce any bias in the statistics of morphologically 
disturbed systems, we have split
the OBEY sample into two absolute magnitude bins using the median value of M$_B=-20.98$ mag. 
We find 59\% of disturbance (filled circles in Figure \ref{tal}) in the subsample with 
M$_B<-20.98$ mag (27 galaxies) and 75\% if we consider
M$_B\geq-20.98$ mag (28 galaxies). Thus, we find a slightly larger percentage of disturbed 
morphologies for the lower luminosity ellipticals. The galaxies in the OBEY survey are less
luminous than the PRGs on average (see Figure \ref{tal1}), and thus, any bias caused by this effect would
result in an enhancement of the number of disturbed morphologies in the OBEY sample relative
to the PRGs at $z<0.2$. However, by looking at Figure \ref{tal} is clear 
that there is no correlation between M$_B$ and the level of disturbance of the galaxies in the OBEY 
sample. Although for the galaxies in the EGS sample the M$_B$ distribution is more similar
to that of the PRGs at $0.2\leq z<0.7$ (0.18 of significance according to the KS test; see Section \ref{egs}), 
we have performed the same test as for the ellipticals in the OBEY survey. By splitting the EGS sample 
into galaxies brighter and fainter than M$_B=-21.25$ mag (i.e. the median value), 
we find 60\% and 50\% of disturbance respectively (53 and 54 galaxies included in each bin respectively). 
Thus, we do not find any significant correlation between the levels of morphological disturbance and the 
luminosity of the elliptical galaxies in either the EGS or OBEY samples.

\subsubsection{Surface brightnesses}
\label{mu}

The main result of \citealt{Ramos11} is that 85\% (78\% if we do not consider galaxies with dust features only) 
of the 2Jy sample of PRGs show peculiar optical morphologies at 
relatively high levels of surface brightness: 1) $\tilde{\mu}_V=24.0~mag~arcsec^{-2}$ and 
$\Delta\mu_{V}=[22.1,~26.2]~mag~arcsec^{-2}$ at $z<0.2$; and 2)  $\tilde{\mu}_V=23.5~mag~arcsec^{-2}$ and 
$\Delta\mu_{V}=[21.3,~25.1]~mag~arcsec^{-2}$ at $0.2\leq z<0.7$.

In Tables \ref{mag} and \ref{mag2} we report apparent surface brightnesses ($\mu_{V}$ in the case of the OBEY survey and
$\mu_{R_C}$ for the EGS sample) for all the secure detections of $tails$, $fans$, $shells$, $bridges$, $amorphous$ $haloes$ and 
$irregular features$ detected in the control sample images. These surface brightnesses have been obtained exactly 
in the same manner as those reported in \citealt{Ramos11}. We first calculated the averaged number of counts of each feature
using small apertures, and then repeated the process, using the same aperture, in several regions of the galaxy on either side of the feature
to subtract the sky and the diffuse host galaxy background. In order to test the robustness 
of this technique, CRA and PB measured the surface brightnesses of the same features for some of the galaxies independently, 
obtaining differences below 0.1 mag.

For the OBEY survey, in Table \ref{mag} we report the $\mu_{V}^{corr}$ values, obtained after correcting $\mu_{V}$
from Galactic extinction \citet{Schlegel98}. For the galaxies in the EGS sample, which are at
redshifts $0.2\leq z<0.7$, the surface brightnesses were K-corrected, using the values reported in \citet{Frei94} and 
\citet{Fukugita95} for elliptical galaxies, in addition to the (1+z)$^4$ cosmological dimming and extinction corrections\footnote{Values of 
E(B-V) from \citet{Schlegel98} were used for each of the four Suprime-Cam images in the EGS, together with the \citet{Cardelli89} extinction law
to derive the corresponding A$_{R_C}$ values.}. 
The final $\mu_{R_c}^{corr}$ values for the EGS sample are shown in Table \ref{mag2}.
We finally transform the $\mu_{R_C}^{corr}$ values into V-band measurements to compare with the OBEY sample and the PRGs by assuming 
colours of elliptical galaxies from \citet{Frei94} and \citet{Fukugita95}. 

\begin{table*}
\centering
\begin{tabular}{lcll}
\hline
\hline
ID  & Morphology & $\mu_{V}$  (mag~arcsec$^{-2}$)  &  $\mu_{V}^{corr}$  (mag~arcsec$^{-2}$)  \\
\hline
NGC 0584       & 2S,B	    & 25.6, 26.9, 25.9  		&    25.5, 26.8, 25.8			\\
NGC 0596       & S,F	    & 25.7, 26.1			&    25.6, 26.0 			\\  
NGC 0720       & 2F	    & 26.0, 26.0			&    25.9, 25.9 				\\	       
NGC 1199       & \dots      & \dots				&    \dots					\\  
NGC 1209       & T,3F,[D]   & 27.7, 25.5, 25.2, 25.8		&    27.6, 25.4, 25.1, 25.7			\\ 
NGC 1399       & \dots      & \dots				&    \dots					\\
NGC 1395       & 3S	    & 25.9, 26.4, 26.1  		&    25.8, 26.3, 26.0				\\ 
NGC 1407       & \dots      & \dots				&    \dots					\\
NGC 2865       & 3S,2T,[D]  & 24.6, 23.7, 24.7, 25.8, 26.0	&    24.3, 23.4, 24.4, 25.5, 25.7		\\
NGC 2974       & S,[D]      & 23.9				&    23.7					\\	    
NGC 2986       & [B]	    & \dots				&    \dots					\\
NGC 3078       & \dots      & \dots				&    \dots					\\
NGC 3258       & [2N]	    & \dots				&    \dots					\\	     
NGC 3268       & S	    & 25.5				&    25.2					\\
NGC 3557B      & 2I	    & 25.1, 25.0			&    24.8, 24.7 				\\
NGC 3557       & F,[S]      & 25.5				&    25.2					\\
NGC 3585       & 2S	    & 26.0, 25.4			&    25.8, 25.2 				\\
NGC 3640       & S,4F	    & 26.5, 26.1, 25.5, 24.9, 23.9	&    26.4, 26.0, 25.4, 24.8, 23.8		\\
NGC 3706       & 2S	    & 23.9, 25.3			&    23.6, 25.0 				\\
NGC 3904       & S	    & 25.2				&    25.0					\\
NGC 3923       & 4S	    & 25.8, 25.3, 24.8, 24.4		&    25.6, 25.1, 24.6, 24.2			\\				  
NGC 3962       & S	    & 25.7				&    25.6					\\		      
NGC 4105       & 2F,T	    & 26.1, 25.7, 25.3  		&    25.9, 25.5, 25.1				\\
NGC 4261       & T,F	    & 26.8, 26.7			&    26.7, 26.6 				\\
NGC 4365       & F	    & 25.9				&    25.8					\\
IC 3370        & F,S,D      & 25.3, 25.3			&    25.0, 25.0 				\\
NGC 4472       & \dots      & \dots				&    \dots					\\   
NGC 4636       & F	    & 26.9				&    26.8					\\
NGC 4645       & \dots      & \dots				&    \dots					\\
NGC 4697       & \dots      & \dots				&    \dots					\\
NGC 4696       & S,D	    & 26.3				&    26.0					\\
NGC 4767       & 2S,[D]     & 23.7, 26.0			&    23.4, 25.7 				\\
NGC 5011       & \dots      & \dots				&    \dots					\\
NGC 5018       & 3T,3S,[D]  & 25.1, 25.7, 27.1, 24.7, 26.1, 24.5&    24.8, 25.4, 26.8, 24.4, 25.8, 24.2 	\\
NGC 5044       & \dots      & \dots				&    \dots					\\		   
NGC 5061       & T,S	    & 26.0, 25.0			&    25.8, 24.8 				\\
NGC 5077       & [S],[D]    & \dots				&    \dots					\\
NGC 5576       & 3T,S	    & 26.7, 25.3, 25.8, 25.9		&    26.6, 25.2, 25.7, 25.8			\\
NGC 5638       & T,S	    & 28.1, 28.3			&    28.0, 28.2 				\\
NGC 5812       & T	    & 27.8				&    27.5					\\
NGC 5813       & \dots      & \dots				&    \dots					\\
NGC 5846       & 3S,2N      & 26.7, 26.6, 27.0  		&    26.5, 26.4, 26.8				\\
NGC 5898       & 3T,D,2N    & 27.3, 26.5, 27.6  		&    26.9, 26.1, 27.2				\\
NGC 5903       & \dots      & \dots				&    \dots					\\	   
IC 4797        & T,I,[D]    & 26.4, 25.8			&    26.2, 25.6 				\\
IC 4889        & F	    & 26.4				&    26.2					\\
NGC 6861       & D	    & \dots				&    \dots					\\
NGC 6868       & \dots      & \dots				&    \dots					\\
NGC 6958       & 3S,[D]     & 26.2, 26.6, 28.1  		&    26.1, 26.5, 28.0				\\
NGC 7029       & \dots      & \dots				&    \dots					\\
NGC 7144       & \dots      & \dots				&    \dots					\\
NGC 7196       & S,[D]      & 27.1				&    27.0					\\
NGC 7192       & S	    & 27.5				&    27.4					\\	   
IC 1459        & 4S	    & 26.8, 26.6, 26.1, 26.5		&    26.7, 26.5, 26.0, 26.4			\\
NGC 7507       & S	    & 28.1				&    28.0					\\
\hline		     			      		 
\end{tabular}						 
\caption{Surface brightness measurements of the detected features in the V-band (Vega system). 
Column 2 lists our morphological classification (same as in Table 1). Apparent  
surface brightnesses and those corrected of galactic extinction (A$_V$) for secure identifications 
of T, F, S, B, A and I are given in Columns 3 and 4, respectively. 
Brackets in Column 2 indicate uncertain identification.}
\label{mag}
\end{table*}						 

\begin{table*}
\centering
\begin{tabular}{ccccll}
\hline
\hline
IRAC ID &  Dimming & Type & Morphology   & $\mu_{R_C}$    &  $\mu_{R_C}^{corr}$    \\
        &    (mag~arcsec$^{-2}$) &   &       & (mag~arcsec$^{-2}$) &  (mag~arcsec$^{-2}$) \\         
\hline
004162  &    1.7  &  E   & \dots	  &  \dots			  &   \dots			\\ 
006612  &    1.2  &  E   & B,F,[D],[T]    &  23.5, 24.2		  	  &   21.9, 22.6		\\
006613  &    1.1  &  E   & B		  &  24.7			  &   23.1			\\
056690  &    1.8  &  E   & [A],[B]	  &  \dots			  &   \dots			\\
060191  &    2.0  &  E   & F		  &  25.1			  &   22.1			\\
060958  &    1.5  &  PD  & T,[A],[B]	  &  26.2			  &   24.2			\\
061249  &    2.2  &  PD  & [T] 	  	  &  \dots			  &   \dots			\\
066105  &    1.8  &  E   & [A] 	  	  &  \dots			  &   \dots			\\
067417  &    1.4  &  E   & \dots	  &  \dots			  &   \dots			\\
072533  &    1.2  &  E   & S		  &  25.2			  &   23.5			\\
073519  &    1.7  &  E   & [A] 	  	  &  \dots			  &   \dots			\\
074777  &    1.5  &  E   & [S] 	  	  &  \dots			  &   \dots			\\
074924  &    1.5  &  E   & \dots	  &  \dots			  &   \dots			\\
077695  &    1.3  &  PD  & T		  &  26.2			  &   24.5			\\
079968  &    2.0  &  E   & F		  &  26.7			  &   23.5			\\
082325  &    1.9  &  E   & [F] 	  	  &  \dots			  &   \dots			\\
083714  &    1.8  &  E   & F		  &  25.8			  &   23.2			\\
088031  &    1.8  &  E   & F		  &  26.8			  &   24.3			\\
090430  &    1.4  &  E   & A,F,[B]	  &  25.1, 26.2		  	  &   23.1, 24.3		\\
092065  &    1.9  &  E   & B		  &  25.3			  &   22.4			\\
092765  &    1.3  &  PD  & [A],[T]	  &  \dots			  &   \dots			\\
093764  &    1.4  &  E   & [S] 	  	  &  \dots			  &   \dots			\\
094231  &    1.5  &  E   & 2F,[T]	  &  25.0, 25.2		  	  &   22.9, 23.1		\\
094966  &    1.6  &  E   & 2T  	  	  &  26.0, 26.4		  	  &   23.7, 24.1		\\
095727  &    1.4  &  E   & F,S 	  	  &  27.2, 25.7		  	  &   25.3, 23.7		\\
099954  &    1.0  &  E   & [T] 	  	  &  \dots			  &   \dots			\\
102757  &    0.9  &  E   & 2S  	  	  &  26.1, 26.2		  	  &   25.0, 25.0		\\
102982  &    2.0  &  E   & F		  &  25.0			  &   21.8			\\
103198  &    1.4  &  E   & 2N,F,S	  &  26.5, 25.4		  	  &   24.6, 23.5		\\
104038  &    1.6  &  E   & B		  &  24.4			  &   22.0			\\
104729  &    2.1  &  PD  & A		  &  25.5			  &   22.1			\\
105193  &    0.9  &  E   & [S] 	  	  &  \dots			  &   \dots			\\
106324  &    1.0  &  E   & [T] 	  	  &  \dots			  &   \dots			\\
106984  &    1.6  &  PD  & A,[I]	  &  26.2			  &   23.9			\\
111427  &    1.2  &  E   & 2N,T,2I	  &  24.5, 26.4, 25.7 	  	  &   22.9, 24.8, 24.1     	\\
112580  &    1.8  &  E   & [B] 	  	  &  \dots			  &   \dots			\\
113088  &    1.7  &  E   & [B] 	  	  &  \dots			  &   \dots			\\
113577  &    2.2  &  PD  & [A] 	  	  &  \dots			  &   \dots			\\
114966  &    2.1  &  PD  & 2T,S           &  25.2, 25.3, 26.2             &   21.9, 22.1, 22.9          \\
115327  &    1.3  &  E   & 2F,[T]	  &  27.3, 27.6		  	  &   25.6, 25.8		\\
115594  &    1.2  &  E   & 2N,T	  	  &  26.5			  &   24.9			\\
118942  &    1.4  &  E   & \dots	  &  \dots			  &   \dots			\\
119696  &    1.8  &  E   & B,F 	  	  &  24.9, 26.2		  	  &   22.3, 23.6		\\
122098  &    0.9  &  PD  & \dots	  &  \dots			  &   \dots			\\
124509  &    1.3  &  PD  & B,2T,F	  &  24.3, 24.1, 27.3, 27.0       &   22.6, 22.4, 25.6, 25.3    \\
125663  &    1.8  &  E   & [F] 	  	  &  \dots			  &   \dots			\\
126918  &    1.7  &  E   & F,[B]	  &  25.8			  &   23.2			\\
127241  &    2.0  &  PD  & \dots	  &  \dots			  &   \dots			\\
127457  &    1.7  &  E   & 2N,A	  	  &  24.8			  &   22.2			\\
128074  &    1.3  &  E   & B,[F]	  &  26.9			  &   25.2			\\
128416  &    2.0  &  PD  & \dots	  &  \dots			  &   \dots			\\
132682  &    1.2  &  E   & \dots	  &  \dots			  &   \dots			\\
135859  &    1.4  &  E   & [I] 	  	  &  \dots			  &   \dots			\\
\hline		     			      		 
\end{tabular}						 
\end{table*}						 

\begin{table*}
\centering
\begin{tabular}{ccccll}
\hline
\hline
IRAC ID &     Dimming & Type & Morphology   & $\mu_{R_C}$    &  $\mu_{Rc}^{corr}$    \\
        &    (mag~arcsec$^{-2}$) &   &       & (mag~arcsec$^{-2}$) &  (mag~arcsec$^{-2}$) \\         
\hline
138794  &   1.8  &  E   & [T]  	 	 &  \dots			  &   \dots			\\
139190  &   1.6  &  E   & ...  	 	 &  \dots			  &   \dots			\\
140456  &   1.1  &  PD  & 2T		 &  24.5, 25.3		  	  &   23.0, 23.8		\\
140758  &   1.6  &  E   & S		 &  26.0			  &   23.8			\\
141714  &   1.6  &  PD  & [B],[S]	 &  \dots			  &   \dots			\\
143149  &   1.4  &  E   & T		 &  25.8			  &   23.9			\\
143536  &   1.8  &  PD  & [T]  	 	 &  \dots			  &   \dots			\\
145098  &   1.2  &  E   & A,T  	 	 &  25.7, 27.2		  	  &   24.0, 25.6		\\
145434  &   1.7  &  PD  & 4T		 &  24.6, 25.7, 25.1, 25.7    	  &   22.2, 23.2, 22.6, 23.2    \\
146298  &   2.0  &  E   & [A]  	       	 &  \dots			  &   \dots			\\
152722  &   1.7  &  PD  & [F]  	 	 &  \dots			  &   \dots			\\
156161  &   1.1  &  E   & T		 &  24.5			  &   23.0			\\
157751  &   1.7  &  E   & ...  	 	 &  \dots			  &   \dots			\\
157878  &   1.6  &  E   & F		 &  26.3			  &   24.0			\\
159123  &   1.9  &  PD  & T		 &  25.7			  &   22.8			\\
159936  &   1.5  &  E   & 2N		 &  \dots			  &   \dots			\\
160442  &   1.7  &  E   & B,A  	 	 &  26.5, 25.6		  	  &   24.1, 23.2		\\
160500  &   1.3  &  PD  & B,2T 	 	 &  26.7, 26.0, 25.4 	  	  &   24.9, 24.3, 23.7      	\\
161724  &   1.3  &  PD  & [F]  	 	 &  \dots			  &   \dots			\\
165265  &   2.2  &  E   & B,T  	 	 &  25.7, 26.2		  	  &   22.0, 22.4		\\
166730  &   1.3  &  E   & S,T  	 	 &  25.8, 27.0		  	  &   23.9, 25.1		\\
169386  &   1.7  &  PD  & ...  	 	 &  \dots			  &   \dots			\\
172474  &   1.8  &  E   & T,B,F	 	 &  26.3, 27.9, 26.6 	  	  &   23.7, 25.3, 23.9       	\\
173901  &   1.2  &  E   & ...  	 	 &  \dots			  &   \dots			\\
175347  &   2.0  &  E   & S,[B]	 	 &  26.4			  &   23.2			\\
175590  &   1.9  &  E   & [A]  	 	 &  \dots			  &   \dots			\\
177990  &   1.0  &  PD  & F,[2N]	 &  26.1			  &   24.8			\\
178118  &   1.6  &  E   & ...  	 	 &  \dots			  &   \dots			\\
178724  &   1.8  &  E   & A		 &  26.6			  &   23.9			\\
178868  &   1.4  &  PD  & F              &  26.1                          &   24.2                      \\
180420  &   1.9  &  E   & 2N,2T,[B]	 &  25.8, 26.2 		  	  &   22.9, 23.4		\\
181402  &   1.4  &  E   & [I],[A]	 &  \dots			  &   \dots			\\
181444  &   1.2  &  E   & 2S,[I]	 &  25.5, 24.8 		  	  &   23.9, 23.3		\\
181736  &   1.7  &  PD  & ...  	 	 &  \dots			  &   \dots			\\
181914  &   1.3  &  PD  & ...  	 	 &  \dots			  &   \dots			\\
182762  &   1.5  &  PD  & [F]  	 	 &  \dots			  &   \dots			\\
183081  &   1.3  &  E   & F,[T]	 	 &  26.8			  &   24.9			\\
183836  &   1.6  &  E   & [S]  	 	 &  \dots			  &   \dots			\\
184041  &   1.8  &  E   & F,S            &  25.4, 26.5                    &   22.6, 23.7                \\
184315  &   1.8  &  PD  & 2N		 &  \dots			  &   \dots			\\
186058  &   1.9  &  PD  & [A]  	 	 &  \dots			  &   \dots			\\
189727  &   2.1  &  PD  & ...  	 	 &  \dots			  &   \dots			\\
190795  &   1.8  &  PD  & T,S  	 	 &  25.2, 26.1		  	  &   22.6, 23.4		\\
193464  &   1.5  &  E   & 2N,F 	 	 &  25.5			  &   23.4			\\
193507  &   1.7  &  E   & 2N,[B]	 &  \dots			  &   \dots			\\
193737  &   1.7  &  E   & ...  	 	 &  \dots			  &   \dots			\\
193974  &   1.5  &  E   & [S]  	 	 &  \dots			  &   \dots			\\
194092  &   1.8  &  E   & [T]  	 	 &  \dots			  &   \dots			\\
196827  &   1.4  &  E   & T		 &  26.6			  &   24.7			\\
198295  &   1.9  &  E   & [S]  	 	 &  \dots			  &   \dots			\\
199503  &   1.7  &  PD  & T		 &  26.8			  &   24.3			\\
202111  &   1.0  &  E   & [S]  	 	 &  \dots			  &   \dots			\\
204161  &   2.1  &  E   & A,[B]	 	 &  26.0			  &   22.6			\\
204944  &   1.1  &  PD  & T,S  	 	 &  26.3			  &   24.9			\\
\hline		     			      		 
\end{tabular}						 
\caption{Surface brightness measurements of the detected features in the R$_c$-band (AB system). 
Column 2 corresponds to the surface brightness dimming from NED, column 3 indicates whether 
a galaxy has been visually classified as an elliptical (E) or as a possible disk (PD), and column 4 lists our morphological classification as 
in Table \ref{data}. Apparent and corrected (including 
galactic extinction, dimming, and k-corrections) surface brightness for secure identifications 
of T, F, S, B, A and I are given in Columns 5 and 6, respectively. Brackets in column 4 indicate uncertain identification.}
\label{mag2}
\end{table*}

We have chosen K-corrections and colour transformations for elliptical galaxies, but some of the features may be produced  
in mergers involving small disk galaxies and/or there can be local star formation taking place in tidal 
features associated with the interaction. In such cases the galaxy colours would be
more similar to those of spiral galaxies. In order to assess the importance of this effect, in \citealt{Ramos11} 
we re-calculated the $\mu_{V}^{corr}$ values of the features by using K-corrections  
and colours of Sbc-type spiral galaxies and confirmed that they did not change significantly ($\sim0.1~mag~arcsec^{-2}$).

\begin{figure*}
\centering
\subfigure[]{
\includegraphics[width=8.6cm]{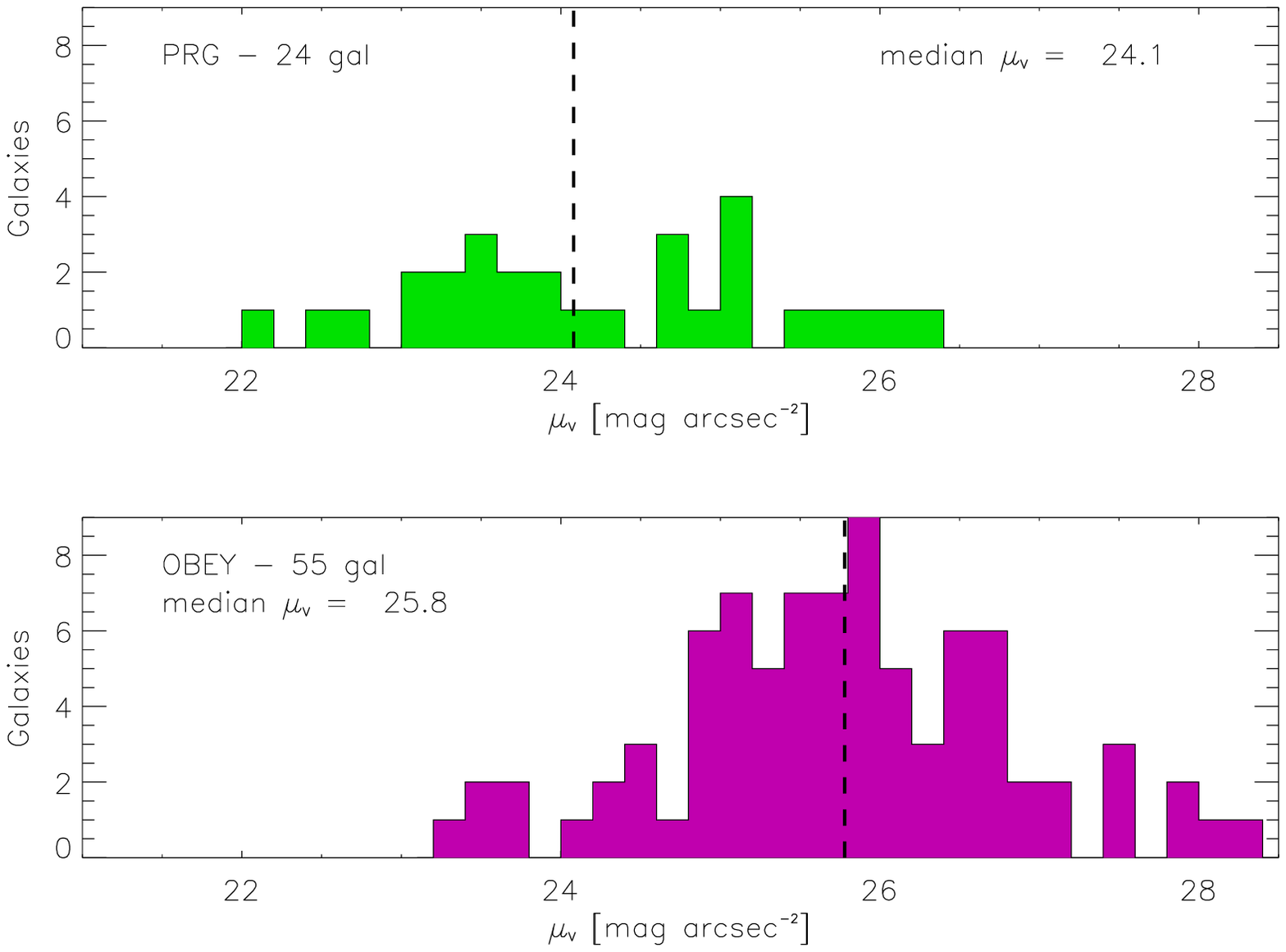}
\label{surf1a}}
\subfigure[]{
\includegraphics[width=8.6cm]{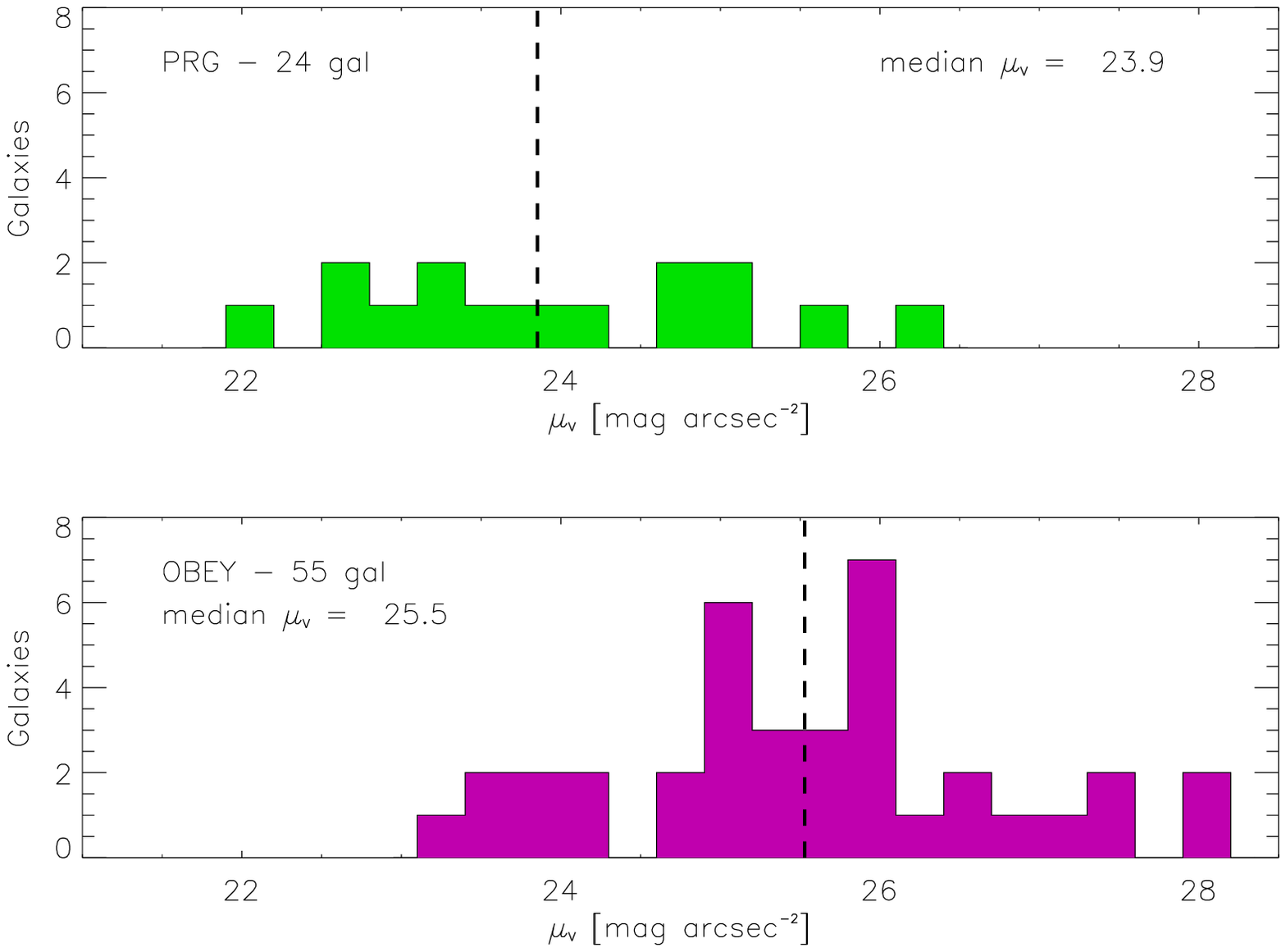}
\label{surf1b}}
\caption{(a) Comparison between the $\mu_{V}^{corr}$ of the 24 PRGs in the 2Jy sample at z$<0.2$ (top panel)
and those of the elliptical galaxies in the OBEY survey (bottom panel). (b) Same as in (a) but considering the
brightest feature in each source only.}
\label{surf1}
\end{figure*}

The comparison between  $\mu_{V}^{corr}$ values
measured for the features detected in our PRGs and those for the OBEY survey is shown in Figure \ref{surf1a} and Table \ref{surf}.
In Figure \ref{surf1b} we show the same comparison, but considering only the brightest disturbed feature of each galaxy.
According to the KS test, the PRG and OBEY distributions shown in Figures \ref{surf1a} and \ref{surf1b} are different 
at the 99.9\% significance level ($>3\sigma$).

The median depth and range of the detected features in the OBEY galaxies are $\tilde{\mu}_{V}^{corr}=25.8~mag~arcsec^{-2}$ and 
$\Delta\mu_{V}=[23.4,~28.2]~mag~arcsec^{-2}$, respectively. {\it Thus, the features detected in PRGs at $z<0.2$
are $\sim$2 mag brighter than in their quiescent counterparts.} 
In fact, if we only consider the features in the OBEY survey which have surface brightnesses within the range of the PRGs at $z<0.2$
(i.e. $\mu_{V}^{corr} \leq 26.2~mag~arcsec^{-2}$, which is exactly the same if we consider SLRGs only; see Table \ref{surf}), 
the percentage of objects with morphological disturbance goes down to 53\%. Thus, when the same range of surface brightness is considered, 
the proportion of interacting systems found in the PRG sample is considerably larger (93\% if we consider SLRGs only) than that 
found for quiescent ellipticals.  

\begin{figure*}
\centering
\subfigure[]{
\includegraphics[width=8.6cm]{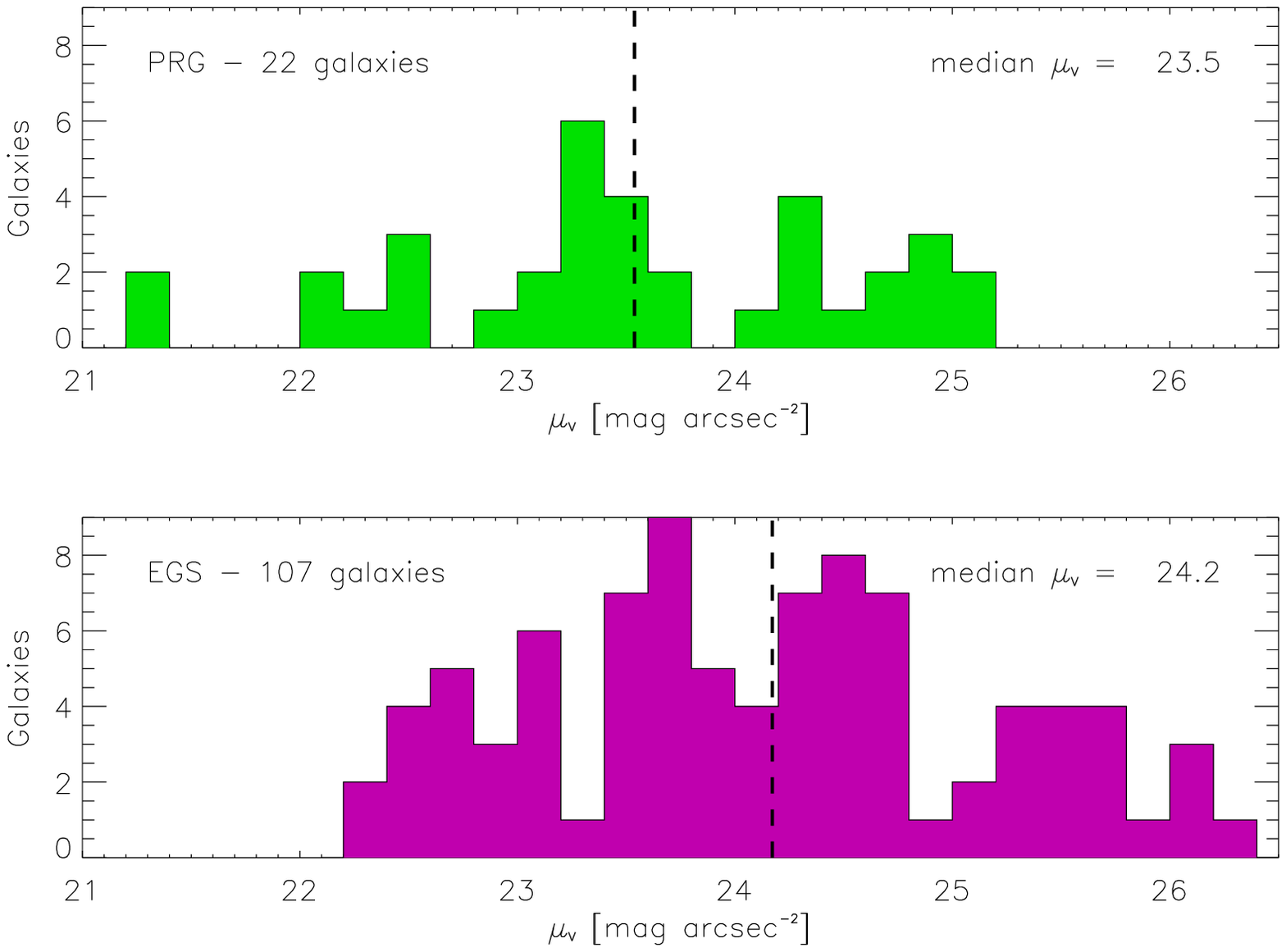}
\label{surf2a}}
\subfigure[]{
\includegraphics[width=8.6cm]{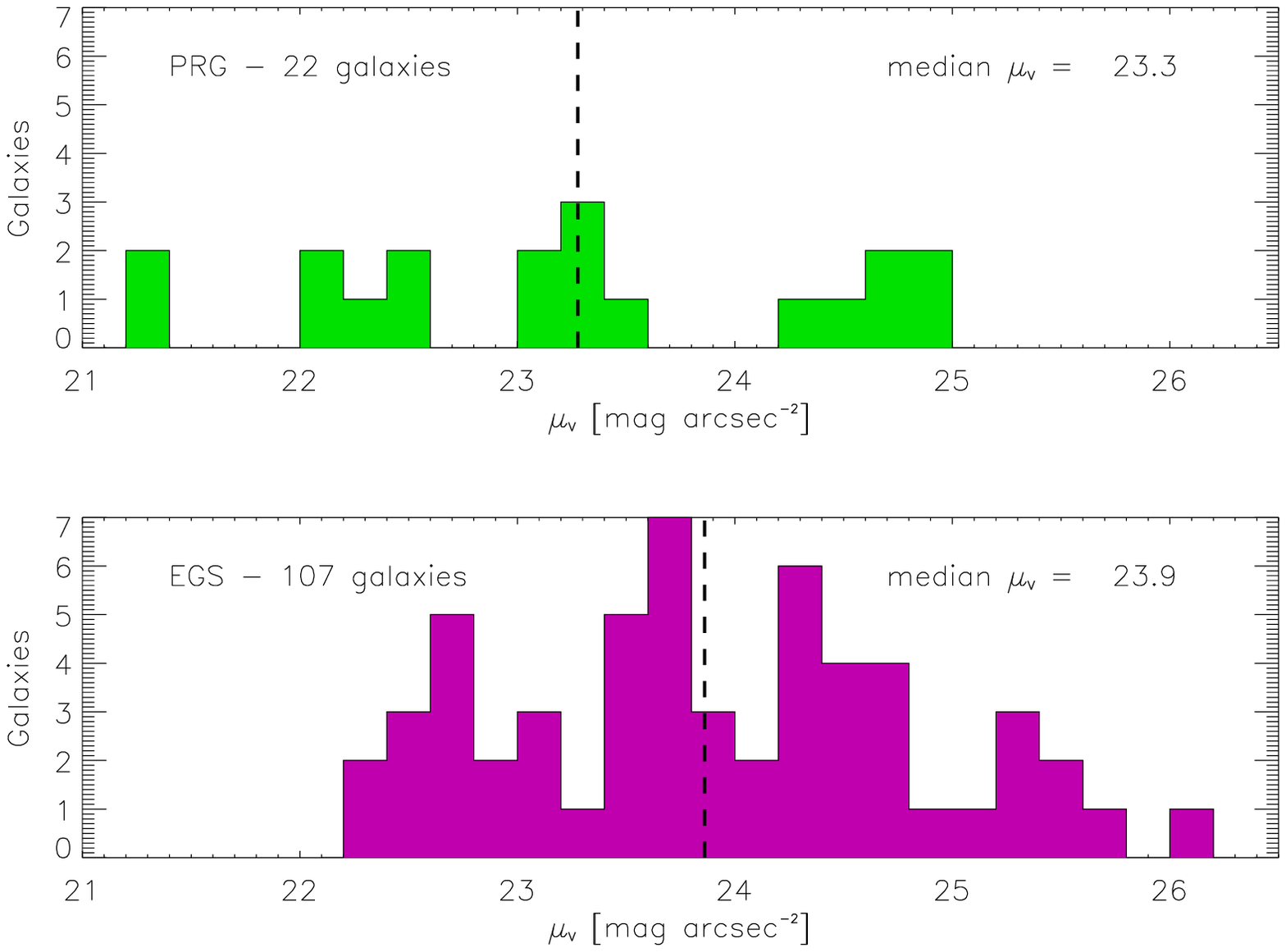}
\label{surf2b}}
\caption{(a) Comparison between the $\mu_{V}^{corr}$ of the 22 PRGs in the 2Jy sample at $0.2\leq z<0.7$ (top panel)
and those of the red galaxies galaxies in the EGS sample (bottom panel). (b) Same as in (a) but considering the
brightest feature in each source only.}
\label{surf2}
\end{figure*}

In Figure \ref{surf2} and Table \ref{surf} we show the same comparisons, 
but this time for the galaxies in the EGS sample and the $0.2\leq z<0.7$ PRGs.
Note that for the galaxies with ``multiple nuclei'' as the only detected feature there are no measurements of 
surface brightnesses and thus, they are not included in Figures \ref{surf1} and \ref{surf2}. However, those 
galaxies have been considered in the statistics presented in Table \ref{same}.
The differences between the two distributions in Figure \ref{surf2a}
are significant at the 2$\sigma$ level: 98\% significance 
according to the KS test and 95\% if we consider the brightest features only 
(Figure \ref{surf2b}).
We measured $\tilde{\mu}_{V}^{corr}=24.2~mag~arcsec^{-2}$ and 
$\Delta\mu_{V}=[22.3,26.3]~mag~arcsec^{-2}$ for the EGS sample. We emphasise that the results obtained by
including or excluding objects with possible disk components (PD) are exactly the same. Thus, for the 
sake of simplicity, in the following we will consider the whole 
sample of 107 galaxies (including ellipticals and possible disks; see Table \ref{data}).

\begin{table*}
\centering
\begin{tabular}{lcccccccc}
\hline
\hline
Morphology & \multicolumn{2}{c}{PRG Sample $z<0.2$} & \multicolumn{2}{c}{OBEY survey} & \multicolumn{2}{c}{PRG Sample $0.2\leq z<0.7$} & 
\multicolumn{2}{c}{EGS sample} \\
  & $\tilde{\mu}_{V}^{corr}$ & $\Delta\mu_{V}$ & $\tilde{\mu}_{V}^{corr}$ & $\Delta\mu_{V}$ & $\tilde{\mu}_{V}^{corr}$ & $\Delta\mu_{V}$ &
$\tilde{\mu}_{V}^{corr}$ & $\Delta\mu_{V}$   \\
\hline
All features 	       & 24.1  &22.1 - 26.2   &25.8 & 23.4 - 28.2& 23.5    &   21.3 - 25.1 & 24.2& 22.3 - 26.3  \\
           	       & (24.3)&(22.6 - 26.2) &\dots& \dots	 & (23.5)  & (21.3 - 25.1) &\dots& \dots	\\			  
Brightest features     & 23.8  &22.1 - 26.2   &25.5 & 23.4 - 28.0& 23.3    &   21.3 - 24.9 & 23.9& 22.3 - 26.1  \\
                       & (24.1)&(22.6 - 26.2) &\dots& \dots	 & (23.3)  & (21.3 - 24.9) &\dots& \dots	\\  
\hline		     			      		 
\end{tabular}						 
\caption{Median values and ranges of the surface brightness measurements of the PRG, OBEY, and EGS samples. Values considering 
all the features detected (two top rows) and the brightest feature of each galaxy (bottom rows) are listed. Surface brightnesses 
between parentheses correspond to SLRGs in the PRG sample.}
\label{surf}
\end{table*}

{\it The surface brightnesses measured for the features of the PRGs at $0.2\leq z<0.7$ are $\sim$1 mag brighter than those
of quiescent red galaxies of similar masses and redshifts.} If we only consider the galaxies in the EGS with features 
within the same $\mu_{V}^{corr}$ range
as the PRGs at $0.2\leq z<0.7$ ($\mu_{V}^{corr}~\leq~25.1~mag~arcsec^{-2}$), 
the percentage of objects with disturbed morphologies is 48\%. Again, this percentage 
is considerably lower than the number of interacting systems found for the PRG sample at $0.2\leq z<0.7$ (95\%; see Table \ref{same}).
In these statistics we have included the systems classified as ``multiple nuclei'', even if it is not
possible to calculate a value of $\mu_{V}^{corr}$ for this type of morphology.

\begin{table*}
\centering
\begin{tabular}{lccccc}
\hline
\hline
Morphology & Group & \multicolumn{1}{c}{PRGs $z<0.2$} & \multicolumn{1}{c}{OBEY} & \multicolumn{1}{c}{PRGs $0.2\leq z<0.7$} & 
\multicolumn{1}{c}{EGS} \\
\hline
Signs of interaction		 & 1,2,3&  62\% (93\%) &  53\%  &  95\% (95\%)  & 48\% (52\%)  \\ 		   
\hline		     			      		 
\end{tabular}						 
\caption{Percentages of disturbance found for the PRGs at $z<0.2$ and the OBEY survey at the same level surface brightness level 
($\mu_V^{corr} \leq 26.2~mag~arcsec^{-2}$) and the same for the PRGs at $0.2\leq z<0.7$ and the EGS survey ($\mu_V^{corr}\leq 25.1~mag~arcsec^{-2}$).
In these numbers we include galaxies classified as multiple nuclei systems (group 3).
Percentages between parentheses correspond to SLRGs in the PRG sample (columns 3 and 5) and to elliptical galaxies only in the EGS sample (column 6).}
\label{same}
\end{table*}

\section{Discussion}

\subsection{Comparison between active and quiescent elliptical galaxies}
\label{comparison}

As explained in Section \ref{intro}, if galaxy interactions are the main triggering mechanism 
for radio-loud AGN activity, then we expect to find stronger and more common signs of morphological disturbance 
in the radio source host galaxies than in the general population of quiescent elliptical galaxies. 
The majority of quiescent luminous ellipticals were likely assembled through gas-poor mergers \citep{vanDokkum05}, 
whereas to trigger and feed a powerful radio source it is likely to require a larger gas supply. According to 
simulations, the morphological 
signatures of gas-rich interactions (such as tidal tails, shells, bridges, etc.) are brighter than those 
produced in gas-poor interactions \citep{Naab06,Bell06,McIntosh08}.

We have compared the morphologies of the 2Jy sample of radio galaxies with first, a sample of 
ellipticals at redshift $z<0.2$, and second, with a sample of early-type galaxies at $0.2\leq z<0.7$. 
We find that a significant fraction of quiescent elliptical galaxies in the two control samples
show evidence for disturbed morphologies at relatively high levels of 
surface brightness, which are likely the result of past or on-going galaxy interactions. However, the morphological 
features detected in the galaxy hosts of the PRGs (e.g. tidal tails, shells, bridges, etc.) are up to 2 magnitudes
brighter than those present in their quiescent counterparts. 

In fact, when we consider the same surface brightness limits for the features in the quiescent galaxies 
and in the PRGs (note that these limits are different in each redshift bin; see Table \ref{same}), we find that the proportion 
of disturbed morphologies in the quiescent population is considerably
smaller (53\% at $z<0.2$ and 48\% at $0.2\leq z<0.7$) than for the PRGs (93\% at $z<0.2$ and 95\% at $0.2\leq z<0.7$, 
considering SLRGs only). This indicates that galaxy interactions are likely to play a role in the triggering of PRG activity. 

However, it is important to recognise that a proportion of the quiescent elliptical galaxy population {\it do} show 
disturbed features at a similar level of surface brightness to the PRGs. Moreover, even if the proportion of such 
objects is relatively small, their volume density  could be considerably larger than that of the (rare) PRGs. This 
raises the question of how the populations of disturbed elliptical galaxies and PRGs are related. The simplest 
assumption we can make is that {\it all} morphologically disturbed ellliptical galaxies go through a radio-loud 
AGN phase at some stage in the galaxy interaction that causes the disturbed features. In this case, the total 
volume density of PRGs ($\rho_{PRG}$) is related to the total volume density of disturbed elliptical galaxies 
($\rho_{DE}$) by the following equation:

\begin{equation}
\frac{\rho_{PRG}}{\rho_{DE}}=0.01\left( \frac{t_{PRG}}{10~Myr} \right) 
\left( \frac{t_{DF}}{1~Gyr} \right)^{-1},
\end{equation}  

where $t_{PRG}$ is the duty cycle of the powerful radio-loud AGN activity and
$t_{T}$ the timescale over which the tidal features associated with a particular galaxy interaction remain 
visible above the surface brightness limit of the observations. Typically, PRGs are expected to remain 
active over a period of $t_{PRG} \sim10 - 100$~Myr \citep{Leahy89,Blundell99,Shabala08}, while the 
tidal features will remain visible on a timescale of $\sim$1~Gyr \citep{Fevre00,Patton02,Conselice03,Kawata06}. 
Therefore we should expect the PRG to make up a fraction $\rho_{PRG}/\rho_{DE}\sim$ 0.01 -- 0.1 of the 
full population of disturbed elliptical galaxies assuming that all such galaxies go through a radio-loud phase.

A direct estimate of the volume density of PRGs can be obtained by integrating the radio luminosity function 
of \citet{Willott01} above the lower radio power limit of the $0.05 < z < 0.7$ 2Jy sample 
($P^{lim}_{151~MHz} \approx 1.3\times10^{25}$~W Hz$^{-1}$ sr$^{-1}$). We find 
$\rho_{PRG}=2\times10^{-7}$~Mpc$^{-3}$ for z=0 and $\rho_{PRG}=1\times10^{-6}$~Mpc$^{-3}$ for 
z=0.5 (corrected to our assumed cosmology). Similarly, an estimate of the volume density of 
disturbed elliptical galaxies can be made by integrating the optical luminosity function for
red sequence galaxies above the lower B-band luminosity limit of the PRG and control samples 
($M_B$=-20.3 mag), then multiplying by the fraction of 
ellipticals with disturbed features of similar surface brightness
to the PRG ($f_D\sim0.5$). For low redshifts we integrate the luminosity function for red sequence 
galaxies from \citet{Baldry04}, obtaining $\rho_{DE}=2\times10^{-4}$~Mpc$^{-3}$, while
for higher redshifts we integrate the $z = 0.5$ luminosity function of \citet{Faber07}, 
obtaining $\rho_{DE}=4\times10^{-4}$~Mpc$^{-3}$. By comparing  
the volume densities of PRGs and disturbed ellipticals estimated in this way we find:  
$\rho_{PRG}/\rho_{DE} \sim 10^{-3}$ for $z < 0.2$ and 
$\rho_{PRG}/\rho_{DE} \sim 2\times10^{-3}$ for $z = 0.5$. These proportions are considerably
lower -- by a factor of five or more -- than those obtained above, based
on the PRG duty cycle and the assumption that all disturbed elliptical galaxies
go through a radio-loud phase ($\rho_{PRG}/\rho_{DE}\sim$ 0.01 -- 0.1). 
{\it We conclude that only a small proportion ($\la$20\%) of 
interacting giant elliptical galaxies with absolute magnitudes $M_B < -20.3$ mag are capable 
of hosting powerful radio sources with radio powers $P^{lim}_{151~MHz} > 1.3\times10^{25}$~W Hz$^{-1}$ sr$^{-1}$
for the requisite timescales.} 

Clearly, while undergoing a galaxy interaction of {\it some type} may be necessary to trigger 
powerful radio jets in a giant elliptical galaxy, it is not by itself sufficient. 
Other potentially important factors include: the degree of gas richness of the interacting galaxies,
their mass ratio, the orbital parameters of the interaction, the masses of the galaxy bulges and 
associated supermassive BHs, and the BHs spin. For example, our classification of morphological disturbance 
in elliptical galaxies is relatively crude and does not precisely distinguish the type of galaxy 
interaction (e.g. whether ``wet'' or ``dry'', minor or major). Therefore it is possible that only 
a minority of the disturbed elliptical galaxies
in our control samples are undergoing the {\it precise type} of interaction that leads to the 
triggering of powerful radio-loud AGN activity.


\begin{table*}
\centering
\begin{tabular}{lccccc}
\hline
\hline
Work & Objects & Sample & Redshift & $\Delta\mu_V$ considered &  Signs of  \\
     &         &        &          & (mag~arcsec$^{-2}$) &      interaction\\    
\hline
\citet{Malin83}     & QE           & 137 & $<$0.01  & $\la$ 25.5        & $\sim$10\%  \\
OBEY survey         & QE           &  55 & $<$0.01  & [23.3, 25.5]     & 34\%        \\
2Jy sample          & SLRGs        &  14 & 0.05-0.2 & [22.1, 25.5]     & 79\%        \\ 
\hline
EGS sample          & ET           & 107 & 0.2-0.7  & [22.3, 25.5]     & 53\% (57\%) \\
2Jy sample          & SLRGs        &  21 & 0.2-0.7  & [21.3, 25.1*]    & 95\%        \\        
\hline
\end{tabular}						 
\caption{Results for the OBEY, EGS and PRG samples considering features with $\mu_V \leq 25.5~mag~arcsec^{-2}$, to 
compare with those found by \citet{Malin83} for a sample of 137 quiescent elliptical galaxies 
(QE) at $z<0.01$. * For the PRGs at $0.2\leq z<0.7$ the dimmest feature detected has 
a $\mu_V$=25.1 mag~arcsec$^{-2}$. In these percentages we include galaxies with double nuclei as the only detected feature (group 3).}
\label{malin}
\end{table*}


Finally, we note that there is an important caveat to bear in mind when making the 
comparison between the PRG and control samples. 
We have matched the comparison samples in galaxy luminosity, redshift, and depth of 
observations (see Section \ref{mu} on the latter), 
but we have not considered the environments of the galaxies. The environment may affect 
the comparison in two ways. First, if a specific type of galaxy interaction is required 
to trigger a PRG (e.g. a major, gas-rich merger), then that type of interaction may be 
favoured by a particular environment (e.g. group rather than cluster; see \citealt{Hopkins08b}). 
Second, the tidal effects
associated with high density environments can rapidly disrupt the morphological structures 
(e.g. shells or ripples; \citealt{Malin83}) that we use to classify the galaxies in our samples.  
Indeed, \citet{Tal09} and \citet{Malin83} explored the relationship between galaxy 
morphology and environment in 
their samples of nearby elliptical galaxies and found that the ellipticals in clusters 
generally appear less disturbed than those in group and field environments. 

Despite the lack of quantification of the galaxy environments in this paper, previous 
studies of radio galaxies in the local Universe have shown that, while FRI sources 
(generally WLRGs) favour clusters, FRII galaxies (generally SLRGs) are found in a wide 
range of environments, ranging from field/group to moderately rich clusters 
\citep{Prestage88,Smith90,Zirbel97}, although there
is some evidence that the environments of FRII objects become richer with
redshift (e.g. \citealt{Hill91}). As noted in Section \ref{obey}, 
the low redshift OBEY control sample 
covers a mix of environments that is similar to that of the FRIIs in the local Universe, 
but less rich on average than that of local FRI galaxies. 
However, our conclusions based on the comparison with the control samples 
would only likely be affected by environmental issues if
the control samples were {\it more} biased towards rich environments than our PRG
sample. At present we have no evidence that this is the case, but the whole
issue of matching control sample environments and the dependence of the degree of 
morphological disturbance on environment clearly warrants further investigation.

\subsection{Evolution of elliptical galaxies from z$\sim$0.7}
\label{ellipticals}

By comparing the galaxies in the OBEY and EGS samples, we can study how the morphologies of elliptical galaxies
evolve from redshift z=0.7 and compare with the predictions of galaxy evolution models. 
Elliptical galaxies are key to investigate the history of galaxy mass assembly, since they dominate the high-end of 
the local luminosity function. While many studies support a scenario in which old (1-4 Gyr) and massive (M$_*>10^{11}M_{\sun}$) 
ellipticals passively evolve from redshift z$\sim$1 (e.g. \citealt{Bundy06} and references therein), 
others argue for a major role of dry-mergers 
in the build-up of the most massive early-type galaxy population from z$\sim$1 
(e.g. \citealt{Bell04,Faber07,Kaviraj07,Lopez11}).

In Table \ref{malin} we show the percentage of disturbance found for the OBEY survey (elliptical galaxies in the 
local universe) and the EGS sample (early-type galaxies with $0.2<z<0.7$) only considering features brighter 
than $\mu_V$=25.5 mag~arcsec$^{-2}$. This value was chosen to match the limiting surface brightness of the features
found by \citet{Malin83} for a sample of 137 elliptical galaxies at $z<0.01$. The latter authors used visual 
inspection of photographic plates to search for shells and ripples, finding that only 10\% of the
ellipticals showed these features. This percentage
is considerably lower than the 34\% that we find for the OBEY survey when the same depth is considered, likely 
due to the limitations associated to the use of photographic images. In addition, in \citet{Malin83} the authors
were looking for sharp, shell-like features with the galaxy at the centre of curvature, rather than more asymmetric
features such as fans, tails, bridges, etc (D. Carter, private communication).

The range of absolute magnitude that we are considering by putting together 
the galaxies in both the OBEY and EGS samples is M$_B$=[-22.5,-20.4] mag (see Section \ref{selection}). 
The percentage of disturbed morphologies in the local 
universe is 34\% and increases to 53\% at z=[0.2, 0.7] when the same depth is considered ($\mu_V\la 25.5~mag~arcsec^{-2}$). 
Thus, we find that a significant
fraction of quiescent elliptical galaxies at low and intermediate redshifts show signatures of past
interactions at relatively high levels of surface brightness, and that this fraction increases slightly with 
redshift. This is consistent with the idea that elliptical galaxies have undergone some evolution 
since z=0.7. However, the interactions that lead to this evolution cannot, in most cases, have noticeably 
modified their star formation 
histories and masses \citep{Lopez11}. This would explain why in the past the most massive elliptical galaxies
were thought to passively evolve from z=1 to z=0.


\section{Conclusions}

We present the results from a comparison between the optical morphologies of a complete sample of 46 southern 
2Jy radio galaxies at intermediate redshifts ($0.05<z<0.7$) and those of quiescent early-type galaxies within
the same mass and redshift ranges. Based on these results, we discuss the role of galaxy interactions in the triggering 
of PRGs. Our major results are as follows:

\begin{itemize}

\item We find that a significant fraction of quiescent early-type galaxies 
across the full redshift range of our study show evidence for disturbed morphologies at relatively high levels of 
surface brightness, which are likely the result of past or on-going galaxy interactions. 

\item The morphological features detected in the galaxy hosts of the 2Jy sample of PRGs 
(e.g. tidal tails, shells, bridges, etc.) are up to 2 magnitudes brighter than those present in their quiescent counterparts.

\item The fraction of disturbed morphologies in the quiescent population is 
considerably smaller (53\% at $z<0.2$ and 48\% at $0.2\leq z<0.7$) than for PRGs (93\% at
$z<0.2$ and 95\% at $0.2\leq z<0.7$, considering SLRGs only) when the same surface brightness 
limits are considered. 

\item These results support a scenario in which PRGs, which are likely triggered by interactions, 
represent a fleeting active phase of a subset of elliptical galaxies that have recently undergone mergers/interactions.

\end{itemize}

\section*{Acknowledgments}

CRA ackowledges financial support from STFC PDRA (ST/G001758/1).
CRA, PGPG, and GB ackowledge 
the Spanish Ministry of Science and Innovation (MICINN) through project Consolider-Ingenio 
2010 Program grant CSD2006-00070: First Science with the GTC (http://www.iac.es/consolider-ingenio-gtc/). 
PGPG and GB acknowledge support from the Spanish Programa Nacional de Astronom\' ia y Astrof\' isica under grants
AYA2009-10368 and AYA2009-07723-E.
KJI is supported through the Emmy Noether programme of the German Science Foundation (DFG).
This work has made use of the Rainbow Cosmological Surveys Database, which is operated by the Universidad Complutense de Madrid (UCM).
This work is based in part on data collected at Subaru Telescope, which is operated by the National Astronomical Observatory of Japan.
This research has made use of the NASA/IPAC Extragalactic Database (NED) which is 
operated by the Jet Propulsion Laboratory, California Institute of Technology, under 
contract with the National Aeronautics and Space Administration.
The authors specially acknowledge Tomer Tal for providing access to the OBEY images, as well as
David Carter, Richard Pogge, Philip Massey, Giovanni Carraro, and Carlos L\'opez-SanJuan for their valuable help.
We finally acknowledge useful comments from the anonymous referee.

\label{lastpage}

\end{document}